%% file: Paper_Heating2018.tex
\documentclass[usenatbib]{mn2e}

\usepackage{amsmath}
\usepackage{natbib}
\usepackage{epsfig}
\usepackage{txfonts}

\input{Befehle}
\usepackage{color}

\usepackage{hyperref}
\usepackage{grffile}
\usepackage{graphics}
\usepackage{booktabs}

\title[Constraints on PMF from magnetic heating]
{Constraints on Primordial Magnetic Fields from their impact on the ionization history with Planck 2018}
\author[D. Paoletti, J. Chluba, F. Finelli, J.~A.~Rubi\~{n}o-Mart\'{\i}n]
{D. Paoletti$^{1,2}$\thanks{E-mail: paoletti@iasfbo.inaf.it},~J. Chluba$^{3}$\thanks{E-mail: jens.chluba@manchester.ac.uk},
~ F. Finelli$^{1,2}$\thanks{E-mail: fabio.finelli@inaf.it} and 
J.~A.~Rubi\~{n}o-Mart\'{\i}n $^{4,5}$\thanks{E-mail: jalberto@iac.es} 
  \\
$^{1}$ INAF/OAS Bologna, Osservatorio di Astrofisica e Scienza dello Spazio, Area della ricerca CNR-INAF, via Gobetti 101, I-40129 Bologna, Italy\\
$^{2}$ INFN, Sezione di Bologna, Via Irnerio 46, I-40126, Bologna, Italy\\
$^{3}$ Jodrell Bank Centre for Astrophysics, Alan Turing Building, University of Manchester,Oxford Road Manchester M13 9PL, UK\\
$^{4}$ Instituto de Astrof\'{\i}sica de Canarias, C/V\'{\i}a L\'{a}ctea s/n, La Laguna, Tenerife, Spain\\
$^{5}$ Dpto. Astrof\'{i}sica, Universidad de La Laguna (ULL), E-38206 La Laguna, Tenerife, Spain
}

\begin{document}

\maketitle

\begin{abstract}
We update and extend our previous CMB anisotropy constraints on primordial magnetic fields through their dissipation by ambipolar diffusion and MHD decaying turbulence effects on the post-recombination ionization history. 
We derive the constraints using the latest {\it Planck} 2018 data release which improves on the E-mode polarization leading to overall tighter constraints with respect to {\it Planck} 2015. 
We also use the low-multipole E-mode polarization likelihood obtained by the \texttt{SROLL2} map making algorithm and we note how it is compatible with larger magnetic field amplitudes than the {\it Planck} 2018 baseline, especially for positive spectral indices. 
The 95 \% CL constraints on the amplitude of the magnetic fields from the combination of the effects is $\sqrt{\langle B^2 \rangle} <0.69 (<0.72)$ nG for {\it Planck} 2018 (\texttt{SROLL2}) by marginalizing on the magnetic spectral index.
We also investigate the impact of a damping scale allowed to vary and the interplay between the magnetic field effects and the lensing amplitude parameter.
\end{abstract}



\section{Introduction}
\label{sec:intro}
Cosmological magnetic fields generated prior to recombination in the early Universe may represent the progenitors of the cosmic magnetism we observe today in the large scale structure (LSS) and whose origin is still one of the open questions in cosmology (for reviews on cosmic magnetism see for example \cite{Vachaspati:2020blt,Subramanian:2018xlb,Durrer:2013pga,Widrow:2011hs,Widrow:2002ud}).
In the future, radio and $\gamma$-ray observations will improve our knowledge of cosmological magnetic fields possibly tightening the lower bounds coming from magnetization of the voids in the LSS from distant extreme blazars \citep{Batista:2021rgm,Barai:2018msb, Tavecchio:2010mk,Neronov:2010gir,Taylor:2011bn,Vovk:2011aa} (and Gamma Ray Burst, see the recent work \cite{Wang:2020vyu} )  and possibly detecting magnetic fields in the filaments connecting the LSS \citep{Govoni981,10.1093/mnrasl/slaa142,Vernstrom:2021hru}. 

The importance of such future observations lies in the fact that whereas in galaxies and clusters the astrophysical generation mechanism, through AGN, stellar dynamos, supernovae etc., is still viable, the generation of cosmological magnetic fields with Mpc coherence lengths in voids and filaments by astrophysical mechanisms is challenging and their presence would strengthen the hypothesis of these fields being relics of primordial magnetic fields (PMFs henceforth)  generated in the early universe.

The generation of such PMFs with suitable amplitudes and coherence lengths requires the existence of peculiar conditions in the early Universe, often correlated with the characteristics of the PMFs generated. In this sense a detection of PMFs signatures and the constraints on their characteristics could be a tell tale of non-standard physics in the Early Universe. Classical examples are conditions as the breaking of conformal invariance required during inflation, or the generation of bubbles if the generation takes place during phase transitions. In particular causal/post-inflationary mechanisms produce PMFs with power spectra bounded to a spectral index equal or greater than two \citep{Caprini:2001nb}.
One of the examples of the interplay between the PMFs and the physics in the early Universe is the possibility of shutting down the allowed window for causally generated magnetic fields using upper bounds from cosmology and lower bounds from Gamma ray astrophysics. This would imply an inflationary generation mechanism for the PMFs which in turn would point to the breaking of conformal invariance during inflation and the relevance of an additional fields coupled with the inflaton. 

Being an additional relativistic species in the cosmological plasma before recombination, PMFs affect the Cosmic Microwave Background (CMB) anisotropies in different ways with signatures visible in both temperature and polarization.
Current constraints on the PMF amplitude estimated on the 1 Mpc scale at the nG level are consistent from both the gravitational effects on cosmological perturbations \citep{Paoletti:2019pdi} (for constraints with older datasets see \citep{Paoletti:2010rx,Shaw:2010ea,Paoletti:2012bb,Planck2013params,Ade:2015cva,Zucca:2016iur}) and non-Gaussianities \citep{Brown:2006wv,Seshadri:2009sy,Caprini:2009vk,Shiraishi:2010yk,Trivedi:2010gi,Shiraishi:2011xvp,Shiraishi:2011fi,Shiraishi:2011dh,Trivedi:2011vt,Shiraishi:2012sn,Shiraishi:2012rm,Shiraishi:2013wua,Trivedi:2013wqa,Ade:2015cva}.
Tighter constraints are provided for extreme configurations as the causal case $n_{\mathrm B}=2$ whose PMF amplitude on the Mpc scale is constrained to the pG level and the almost-scale invariant case $n_{\mathrm B}=-2.9$ whose upper bound is set at 2 nG. The future B-mode polarization measurements from {\it LiteBIRD}\citep{LiteBIRD:2022cnt} and ground based experiments should shrink the constraints below the nG level and sub-pG for the causal case \citep{Paoletti:2019pdi}.
Additional constraints can be provided by the Faraday rotation effects which are currently on the $\mu$G level due to the lack of low frequency high sensitivity polarization maps \citep{Ade:2015cva, Kahniashvili:2008hx} but the situation will change with the next generation of experiments with rotational angle estimators promising to break the nG level also for this effect \citep{Gruppuso:2020kfy,Pogosian:2018vfr}.  
Recently the presence of PMFs magnetic fields at recombination has been also associated to the development of fluctuations in the recombination itself that may alleviate the Hubble tension \citep{Jedamzik:2020krr}.

One of the main effects on CMB anisotropies by PMFs is caused by the dissipation of the fields themselves around and after recombination. The dissipated energy is injected in the cosmological plasma increasing its temperature and modifying the ionization history. The two main dissipation mechanisms are the ambipolar diffusion and magnetohydrodynamic (MHD) decaying turbulence which we will analyse and use to constrain the PMFs characteristics with the most recent {\it Planck} 2018 data \citep{Planck:2018nkj}.
We limit our analysis to the effects on the CMB anisotropies and we do not consider the most direct effect of this energy injection into 
the CMB absolute spectrum \footnote{Different from the cyclotron effect derived in \cite{Burigana:2006ic}} \citep{2000PhRvL..85..700J,Kunze2014, Wagstaff:2015jaa}, whose estimate is below the COBE-FIRAS sensitivities \citep{Fixsen:1996nj} and will need to wait for the next generation of CMB spectral distortions measurements  \citep{Delabrouille:2019thj,2021ExA....51.1515C}.

A full rendition of the turbulence effects would have to consider a full simulation approach to account for the transfer of energy at different scales given by turbulence \citep{Durrer:2013pga} that in case PMFs couple with the kinetic component of the plasma can lead to modifications of the time evolution of the PMFs deviating the spectrum of the fields from the simple power law usually assumed \citep{Christensson:2000sp,Christensson:2002xu,Saveliev:2013uva,Kahniashvili:2016bkp,Brandenburg:2016odr}. 
Such a simulation kind of approach is not feasible for our target both in terms of volume required to reach the largest scales necessary for the CMB analysis and in terms of number of PMFs configurations. For a likelihood based analysis we require the PMFs characteristics to be sampled on a range of priors which will by default require a semi-analytical treatment associated with a Boltzmann and Markov Chain MonteCarlo code, therefore, we adopt a simplified treatment based on analytical forms for the energy injections rates \citep{Sethi2005, 2005PhRvD..72b3004S,Sethi2009} that phenomenologically reproduce the impact on the temperature and ionization 
fraction of the MHD decaying turbulence and ambipolar diffusion.  
We will limit our analysis to non-helical fields although helicity plays a crucial role in the turbulence influencing the spectral time evolution of the PMFs, simulations seems also to indicate that PMFs rapidly reach the maximal helical condition independentely from the initial conditions \citep{Christensson:2000sp,Christensson:2002xu,Saveliev:2013uva,Kahniashvili:2016bkp,Brandenburg:2016odr,2018MNRAS.481.3401T}. 

The scope of this work is to update to {\it Planck} 2018 data the previous treatments \citep{Kunze2014, 2015JCAP...06..027K, 2015MNRAS.451.2244C, Ade:2015cva, Paoletti:2018uic} and in particular estimate the impact of the new low-$\ell$ polarization EE mode testing the different products which have been publicly delivered.
We use the same approach as \citep{Paoletti:2018uic,2015MNRAS.451.2244C,Ade:2015cva} with the numerical improvements we developed. 

The paper is organized as follows: in section 2 we introduce the PMFs model and notation; in section 3 we present the ambipolar diffusion and MHD decaying turbulence effects on the CMB angular power spectra; in section 4 we present the constraints on PMFs for different models and data choices; finally in section 5 we draw our conclusions.

\section{Primordial magnetic fields model}
We model the PMFs as a stochastic background which ensures a local generation mechanisms and the standard Friedmann-Lemaitre-Robertson-Walker metric for the background. We characterise the scale dependence of the fields with a power law power spectrum where as usual the two characteristic parameters describing the configuration of the fields are the amplitude and the tilt. We assume non-helical fields  (for a description of the impact of helicity in the PMFs dissipation see \cite{Jagannathan:2020qob}).
The generic two point correlation function for a stochastic background is given by:
\begin{equation}
\langle B_i({\bf k}) B_j^*({\bf k}')\rangle=(2\pi)^3 \delta({\bf k}-{\bf k}')
 (\delta_{ij}-\hat k_i\hat k_j) \frac{P_{\mathrm B}(k)}{2}
 \end{equation}
with the power spectrum being the usual power law $P_{\mathrm B}(k)=A_{\mathrm B} k^{n_{\rm B}}$. 

We express the amplitude of the power spectrum with the amplitude of the magnetic fields themselves:
\begin{equation}
\langle B^2 \rangle = \frac{A_{\mathrm B}}{2 \pi^2} \int_0^{k_{\mathrm D}} dk k^{2+n_{\mathrm B}} = \frac{A_{\mathrm B}}{2 \pi^2 (n_{\rm B}+3)} k_{\rm D}^{n_{\rm B}+3}\,.
\label{rms_sharp}
\end{equation}
The scales relevant for this work are the observational window of CMB which encompasses large and intermediate scales. Within this observational region we can assume an ideal MHD limit and neglect non-linear effects and possible backreactions of the fluid onto the PMFs which would lead to a different time-evolution of the fields \citep{Saveliev:2012ea,Saveliev:2013uva,Brandenburg:2016odr}.
In these limits the temporal evolution of the PMFs is decoupled from the spatial one and the fields are simply diluted by the Universe expansion as a radiation-like component $\rho_{\mathrm{B}}({\bf{x}},\tau)=\frac{\rho_{\mathrm{B}}({\bf{x}})}{a^4(\tau)}$ with $B({\bf{x}},\tau)=\frac{B({\bf{x}})}{a^2(\tau)}$. 

Magnetically induced perturbations and in general magnetosonic waves are suppressed on small scales by radiation viscosity but due to their intrinsic nature their damping scale is smaller than the classic Silk damping scale for acoustic waves in cosmological perturbations. 
We assume a damping scale which is dependent on both the cosmological background and the field configuration in order to account for the effects of different scale-power distribution on the damping, we use the scale from  \citep{Jedamzik1998, Subramanian1998}:
\begin{equation}
{\bf \bf k_{\rm D} = 
\frac{\sqrt{5.5\times 10^4} (2\pi)^\frac{n_{\rm B}+3}{2}}{\sqrt{\langle  B^2 \rangle} /nG \sqrt{\Gamma[(n_{\rm B}+5)/2]}}\, \sqrt{h \,\frac{\Omega_{\mathrm B} h^2}{0.022}}{\bf \Mpc^{-1}}}\,.
\label{KD}
\end{equation}
The dependence on the physics of the damping scale is an important issue we previously flag \cite{Paoletti:2018uic},
as the results depend on the model and scale of the damping assumed.  Further discussion on this issue can be found in \cite{2018MNRAS.481.3401T}.

\section{Post-Recombination dissipation of the magnetic fields}

The main trigger for the dissipation effects to kick in during and after recombination is the dropping of the ionization fraction. When the ionization fraction drops below the level of 1 ion over 10000 neutral atoms, the velocity difference between the ionized and neutral components lead to the development of ambipolar diffusion and the reduced coupling between baryon and photons lead to the development of MHD decaying turbulence.

Both the effects dissipate magnetic energy into the cosmological plasma causing an increase in temperature. This increase can be parametrized in terms of an injection rate of energy contribution to the electron temperature \citep{Sethi2005}:
\begin{equation}
\label{eq:dT_dt}
\frac{\id \Te}{\id t}=- 2 H \Te 
+ \frac{8 \sigT\Ne \,\rho_\gamma}{3\me c N_{\rm tot}} (\Tg-\Te) + 
\frac{\Gamma}{(3/2)k N_{\rm tot}} \,,
\end{equation}
where $\rho_\gamma=a_{\rm R} \Tg^4\approx 0.26 \, {\rm eV} (1+z)^4$ is the CMB energy density, $N_{\rm tot} =N_{\rm H}(1+f_{\rm He}+X_{\rm e})$ is the number density of matter particles, $N_{\rm H}$ is the number density of hydrogen nuclei, $f_{\rm He}\approx \Yp/ 4(1-\Yp)$ that for $\Yp=0.24$ is $\approx 0.079$,$X_{\rm e}=N_{\rm e}/N_{\rm H}$ is the free electron fraction and finally $H(z)$ is Hubble rate.
The three terms in the equation describe: the heating due to the PMF effect in the third place; the Compton term which accounts for both heating and cooling depending on the relative temperatures of the electron and photon components in second place; and the simple cooling due to the dilution by the Universe expansion, parametrized by the Hubble constant, in the first place. 
We can encapsulate the two dissipation effects in terms of two different injection rates.

\subsection{Ambipolar diffusion}

Ambipolar diffusion is caused by the different velocities of the ionized and neutral components in partially-ionized magnetized plasmas. The collisions between the two components thermalize the plasma by transferring the dissipated energy to the neutral component heating the plasma. 
The injection rate can be approximated by \citep{Sethi2005, Schleicher2008b}:
\begin{align}
\Gamma_{\rm am}\approx \frac{(1-X_{\rm p})}{\gamma X_{\rm p}\, \rho_{\rm b}^2} \left< {\bf L}^2\right>
\end{align}
where $\left< {\bf L}^2\right>$ is the Lorentz force, $\rho_{\rm b}$ is the baryon density,  $X_{\rm p}$ is the coupling between the ionized and neutral components. 
We assume the same numerical treatment we implemented in our previous work \citep{Paoletti:2018uic} where the numerical integration of {\tt Recfast++} \citep{2011MNRAS.412..748C} was improved in order to be able to perform the integration also with positive spectral indices PMFs maintaining the stability.
\subsection{Decaying MHD turbulence}
After recombination, the radiation viscosity decreases and the magnetized fluid is not anymore able to maintain low Reynolds number causing the development of MHD decaying turbulence on scales smaller than the Jeans length. 
The rate for MHD decaying turbulence can be approximated by \citep{Sethi2009}:
\begin{align}
 \Gamma_{\rm turb}=\frac{3 m}{2}\, 
 \frac{\left[\ln \left(1+\frac{t_i}{t_{\rm d}}\right)\right]^m}{\left[\ln \left(1+\frac{t_i}{t_{\rm d}}\right)
 + \frac{3}{2} \ln \left( \frac{1+z_i}{1+z}\right)\right]^{m+1}} H(z)\,\rho_{\rm B}(z),
 \label{Eqn:rate}
 \end{align}
with $t_i/t_{\rm d}\approx 14.8 (\langle  B^2 \rangle^{1/2} / {\rm nG})^{-1}(k_{\rm D}/ \Mpc^{-1})^{-1}$,$m=2(n_{\rm B}+3)/(n_{\rm B}+5)$ and $\rho_{\rm B}(z)=\langle  B^2 \rangle (1+z)^4/ (8\pi) \approx \pot{9.5}{-8} (\langle  B^2 \rangle/ {\rm nG}^{2})\,\rho_\gamma(z)$.
As in our previous work we introduce a Gaussian smoothing for the flaring up of the turbulence at recombination to avoid cusps in the derivative of the rate which may cause numerical instabilities. This is also justified by a slower raise in the onset of MHD turbulence from recent 3D simulations \citep{2018MNRAS.481.3401T}. The rate is then modelled as:
\begin{itemize}
 \item for $z<z_i\sim1088$, Eq.~\eqref{Eqn:rate};
 \item for $z_i \le z \le 1.001 z_i$ polynomial to smooth the derivative at $z_i$ and make it zero at $1.001 z_i$;
 \item for $z>1.001 z_i$ Gaussian suppression to model the onset of turbulent heating.
 \end{itemize}
 Due to the temporal narrowness of this smoothing solution which is restricted around recombination time, it does not affect the global effect of MHD decaying turbulence on CMB anisotropies as demonstrated in our previous work \cite{Paoletti:2019pdi}.  
  \begin{figure}
 \centering
 \includegraphics[width=1\columnwidth]{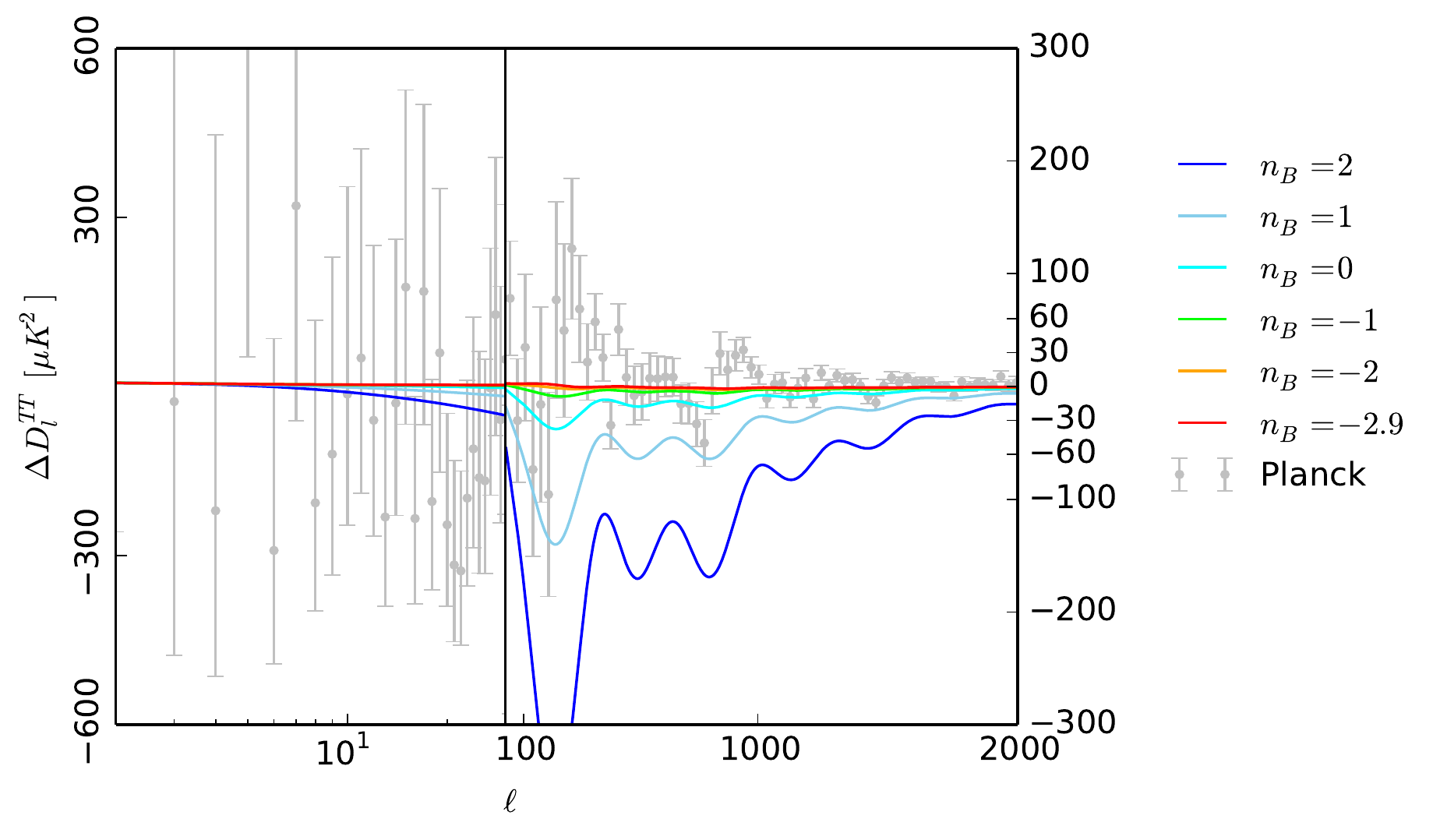}\\
 \includegraphics[width=1\columnwidth]{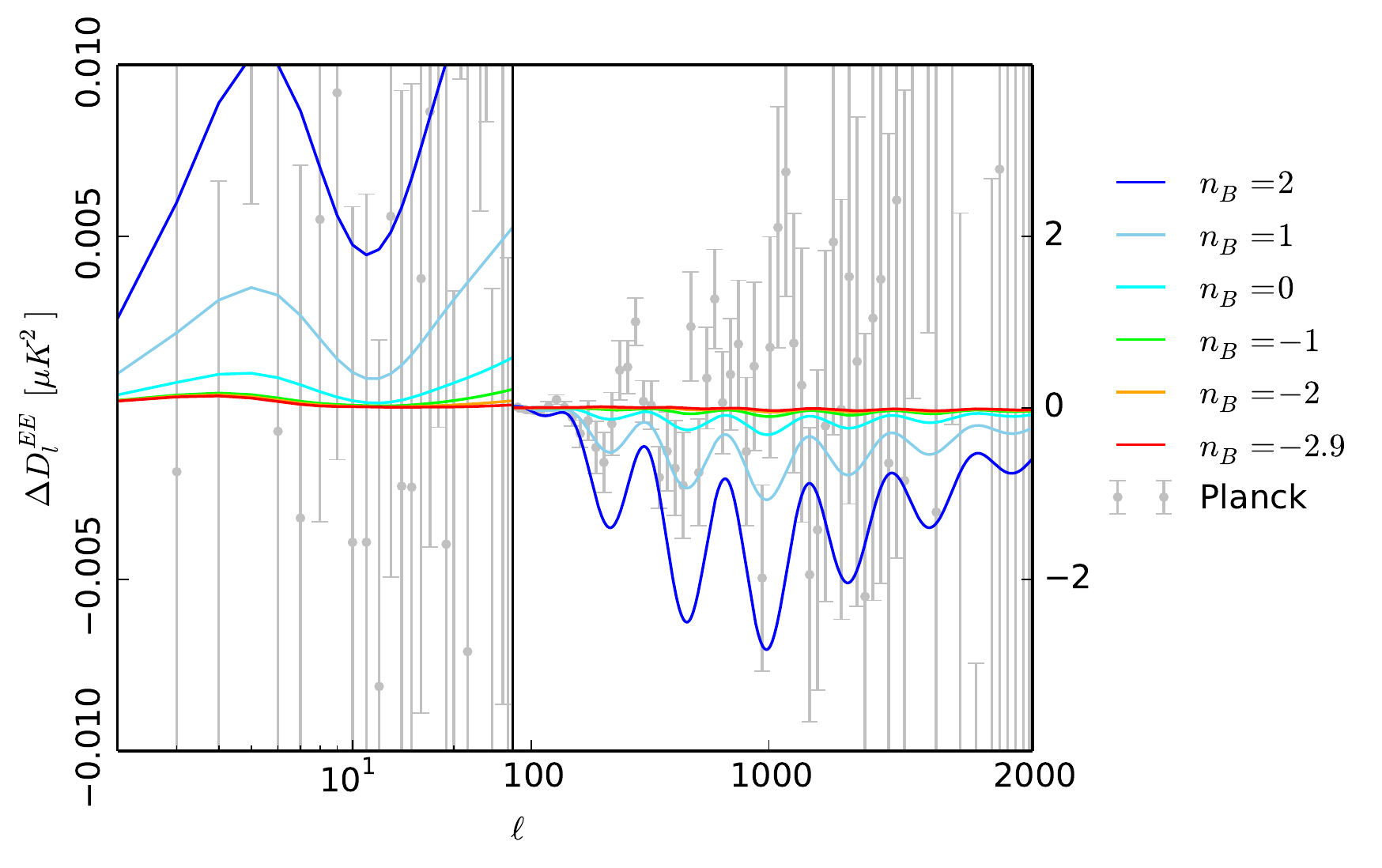}
 \caption{Effect on the CMB angular power spectrum in temperature and polarization of the ambipolar diffusion. We present the differences with respect to the {\it Planck} 2018 best fit with the {\it Planck} 2018 points in silver colour. Coloured lines represent the different spectral indices from colder to warmer colours follow higher to lower spectral indices.Here for graphical reason the amplitude of the PMFs is fixed to 0.1 nG.}
\label{fig:AMBI_APS}
 \end{figure}
\subsection{CMB angular power spectra}
Both ambipolar diffusion and MHD decaying turbulence affect CMB anisotropies in temperature and polarization, but in slightly different ways which make them more or less dominant depending on the configurations of the magnetic fields. 
In \autoref{fig:AMBI_APS} we review the effects of ambipolar diffusion into the CMB angular power spectra in temperature and E-mode polarization. 
The effect is presented relatively to the {\it Planck} 2018 data baseline best fit and compared with the actual data points in gray. Due to the strength of the effect we chose a relatively low amplitude of 0.1\,nG in order to show the variation with the spectral index and the comparison with data points in a clear way in a reasonable plot range. For ambipolar diffusion the effect is strongly dependent on the configuration of PMFs, with configurations which are tilted in a way to give more power to the smaller scales, the blue spectral indices, having stronger impacts on both temperature and polarization. Going towards the scale invariance instead reduces the effect. The effect in temperature anisotropies are mainly in the region of the acoustic peaks, as expected due to the impact on the ionization history, whereas in polarization there is also a strong effect in the region of the reionization bump again expected for the increase of temperature and as a consequence of the optical depth.
 \begin{figure}
 \centering
 \includegraphics[width=1\columnwidth]{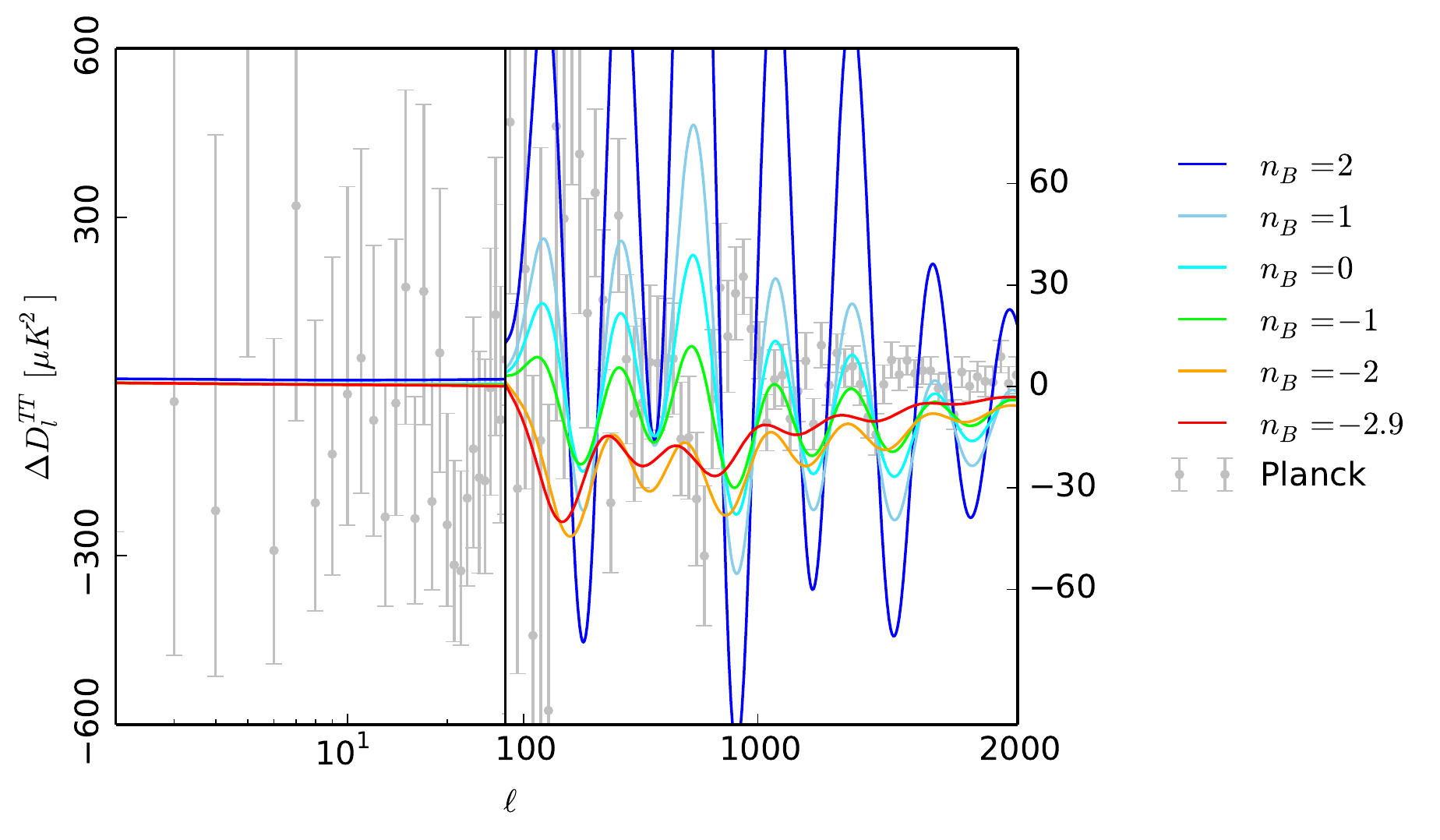}\\
 \includegraphics[width=1\columnwidth]{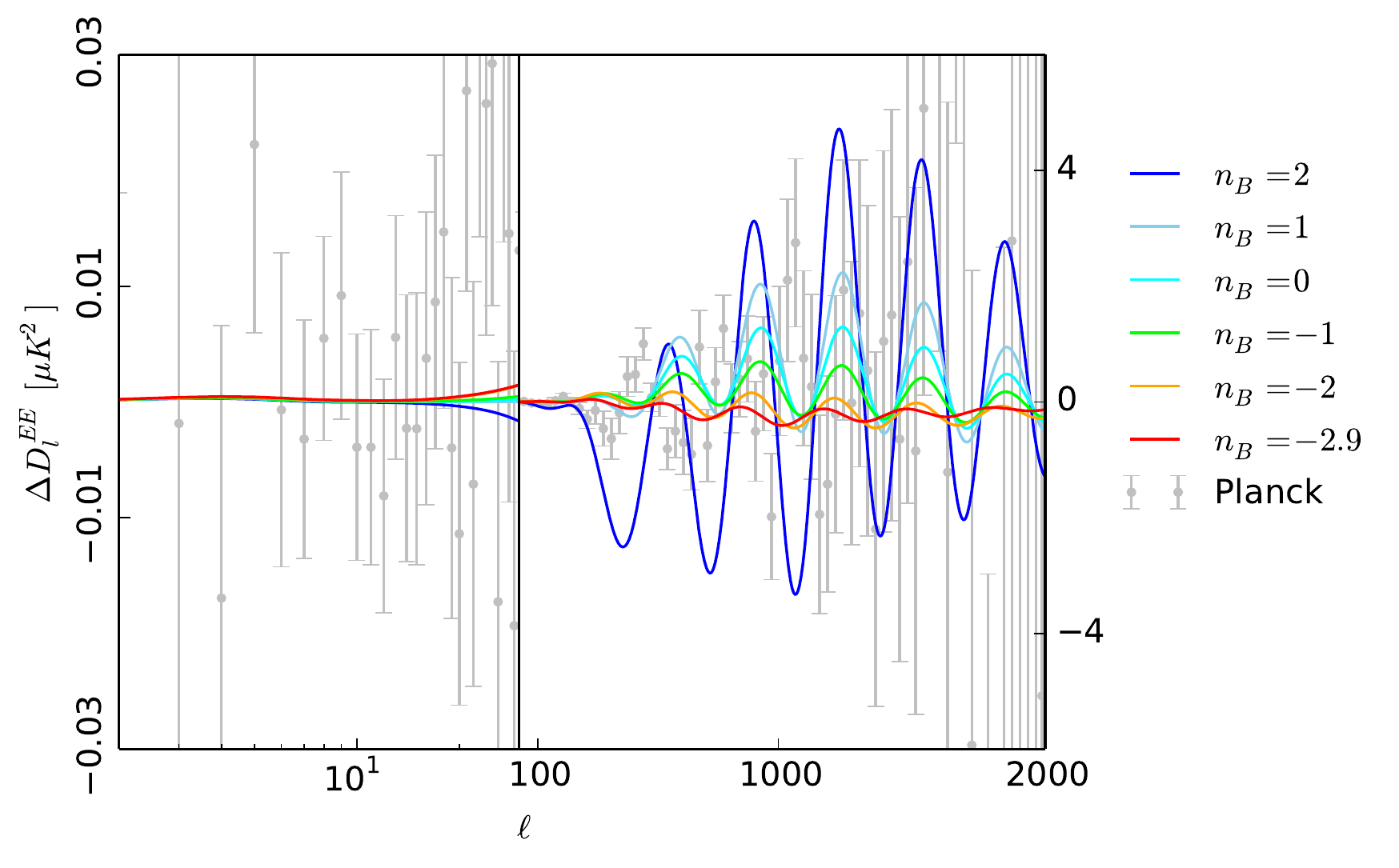}
 \caption{Effect on the CMB angular power spectrum in temperature and polarization of the MHD turbulence. We present the differences with respect to the {\it Planck} 2018 best fit with the {\it Planck} 2018 points in silver colour. Coloured lines represent the different spectral indices from colder to warmer colours follow higher to lower spectral indices.The amplitude is 1 nG.}
\label{fig:MHD_APS}
 \end{figure}
 
 In \autoref{fig:MHD_APS} we review the effects of the MHD decaying turbulence again relatively to the {\it Planck} 2018 best fit and compared with the data points. In this case we choose 1 \,nG as the amplitude of the PMF to evidence the effect.
  As for the ambipolar diffusion case in temperature the main effect is limited to the acoustic peaks region but in this case we observe a modulation of the oscillation effect rather than an overall increase or decrease. Due to its more integrated nature the MHD decaying turbulence has a different effect in the E-mode polarization with minimal if no effect at all on large angular scales. The MHD turbulence develops on the smallest scales and therefore its effect is mostly limited to the acoustic peaks region.
 In the MHD case we note a tendency which also is reflected in the PMFs amplitude constraints: there is a much weaker dependence on the spectral index $n_B$ compared to the ambipolar diffusion. 

\begin{figure}
 \centering
 \includegraphics[width=1\columnwidth]{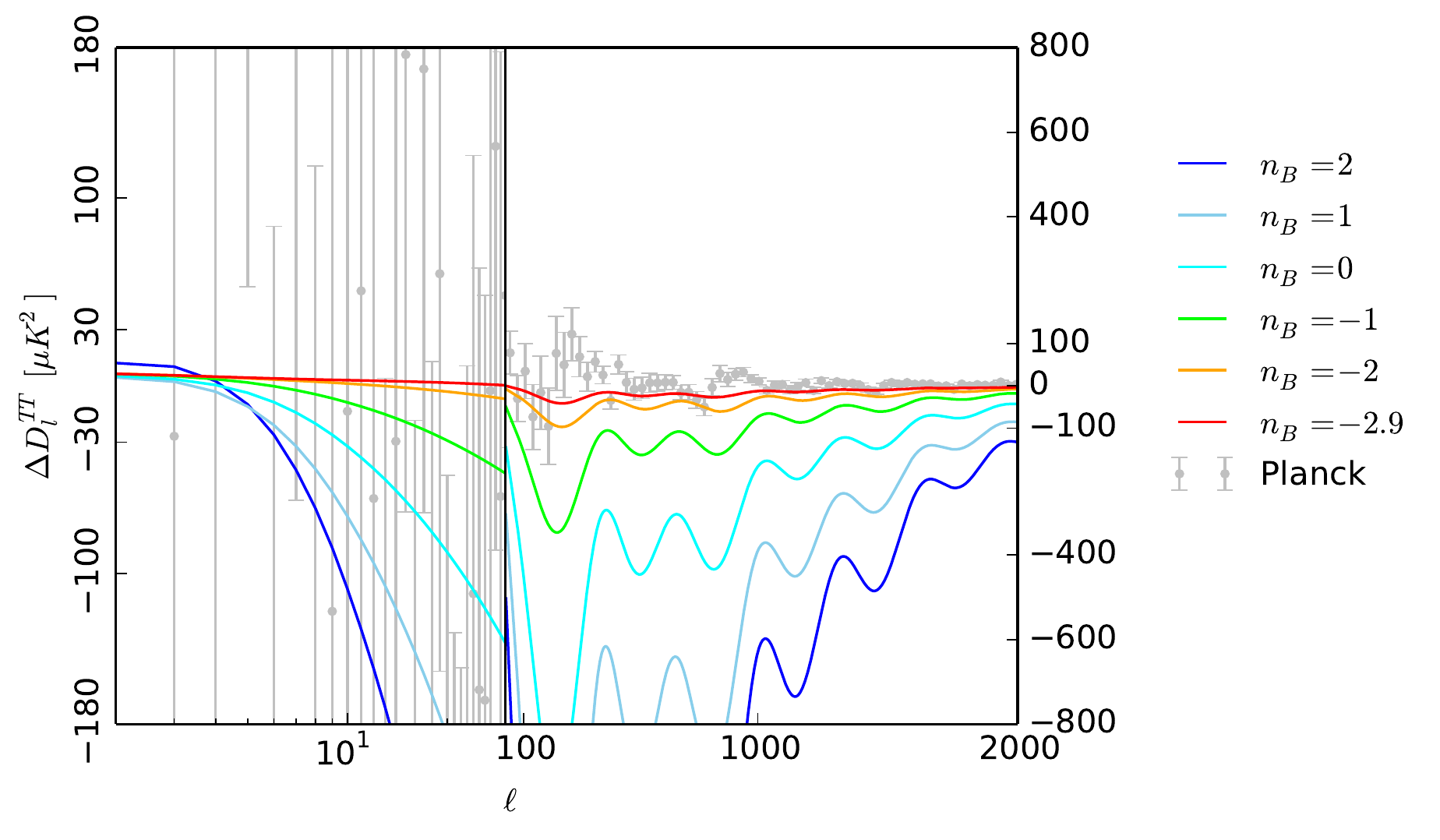}\\
 \includegraphics[width=1\columnwidth]{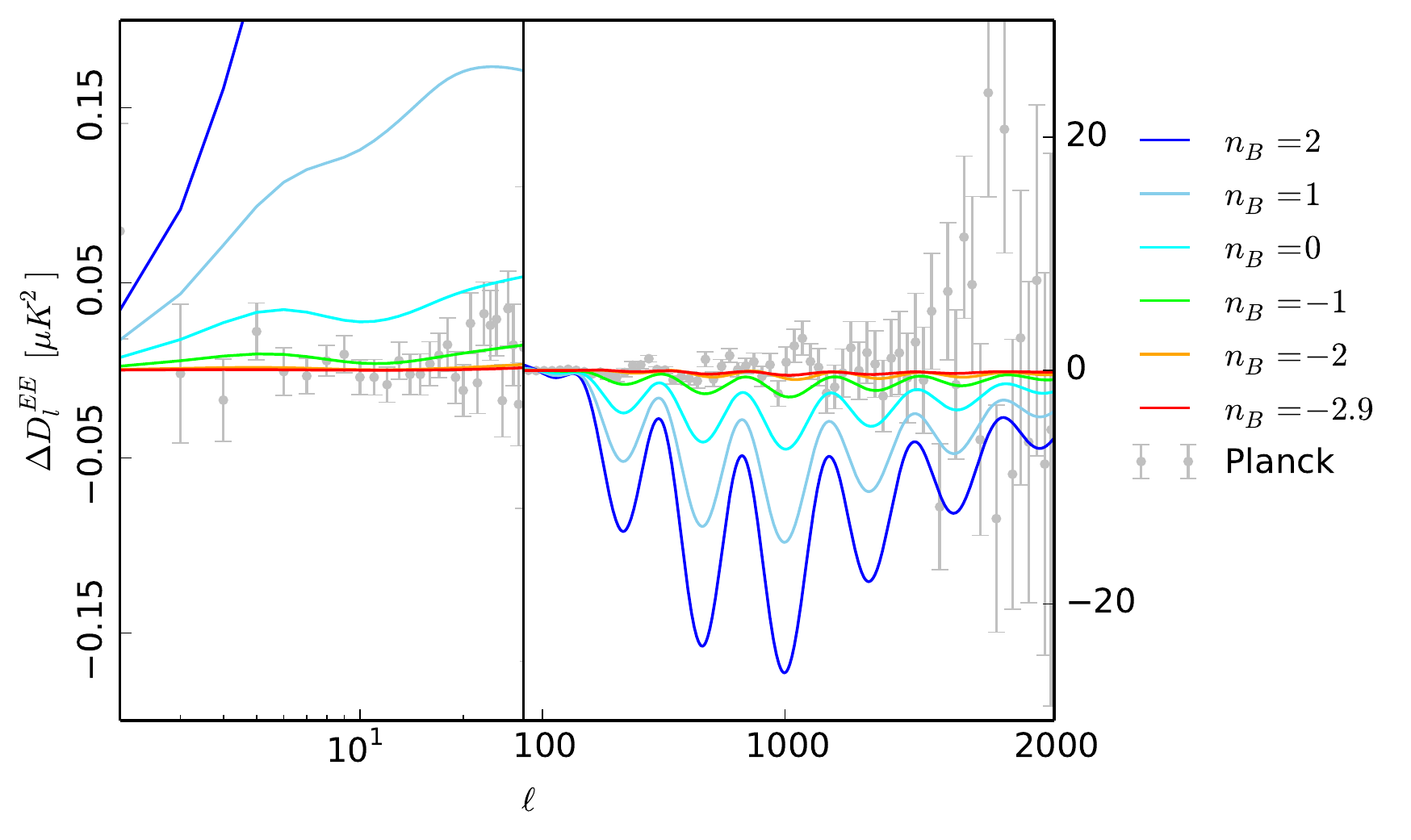}
 \caption{Effect on the CMB angular power spectrum in temperature and polarization of jointly ambipolar diffusion and MHD turbulence. We present the differences with respect to the {\it Planck} 2018 best fit with the {\it Planck} 2018 points in silver colour. Coloured lines represent the different spectral indices from colder to warmer colours follow higher to lower spectral indices.The amplitude of the fields is 1\,nG.}
\label{fig:MHDAMBI_APS}
 \end{figure}

Finally in \autoref{fig:MHDAMBI_APS} we present the combined results for 1\,nG PMFs. The combination of the two effects produces large variations on both large and small angular scales affecting both temperature and polarization. As previously mentioned the effect of ambipolar diffusion for such and amplitude and blue indices is very strong and dominates most of the variation for these kind of PMFs.

\vspace{-3mm}
\section{CMB constraints on the amplitude of PMF}
We derive the constraints on PMF amplitudes for the different configurations by using the {\it Planck} 2018 latest data release. 
{\it Planck} 2018 data improves on the 2015 especially in E-mode polarization where the use of the high frequency cross spectrum $100\times 143$\,GHz has drastically increased the accuracy of the data. Unless otherwise stated we use the {\it Planck} 2018 baseline combination of large scale temperature likelihood based on component separated map, the simulation based {\it \texttt{simall}} likelihood for the E-mode polarization and the high-ell \texttt{plik} likehood for temperature and polarization \citep{Planck:2019nip}. We also add the lensing likelihood based on the extraction of the four point function from temperature maps \citep{Planck:2018lbu}.
We also perform some of the analysis using the update on {\it Planck} 2018 data processing for E-modes on large angular scales of \texttt{SROLL2} \citep{Delouis:2019bub}. \texttt{SROLL2} implements a slightly different map making algorithm which reduces the contamination in the data producing cleaner maps. The associated likelihood has shown to reduce the uncertainty on the value of the optical depth \citep{Pagano:2019tci}

Together with the PMF parameters we vary all the cosmological parameters of the standard model: the baryon density ($\Omega_{\mathrm B} h^2$), the dark matter density ($\Omega_c h^2$), the optical depth ($\tau$), the angular diameter distance to the last scattering surface ($\theta$) and finally the two primordial power spectrum parameters, the scalar spectral index ($n_s$) and its amplitude ($A_s$). We also vary all the nuisance parameters required by the {\it Planck} likelihoods for the foreground and calibration uncertainties. 
Concerning the PMFs parameters we either marginalize over the magnetic spectral index or we fix $n_B$ to relevant physical cases.

\begin{figure}
    \centering
    \includegraphics[width=0.5\textwidth]{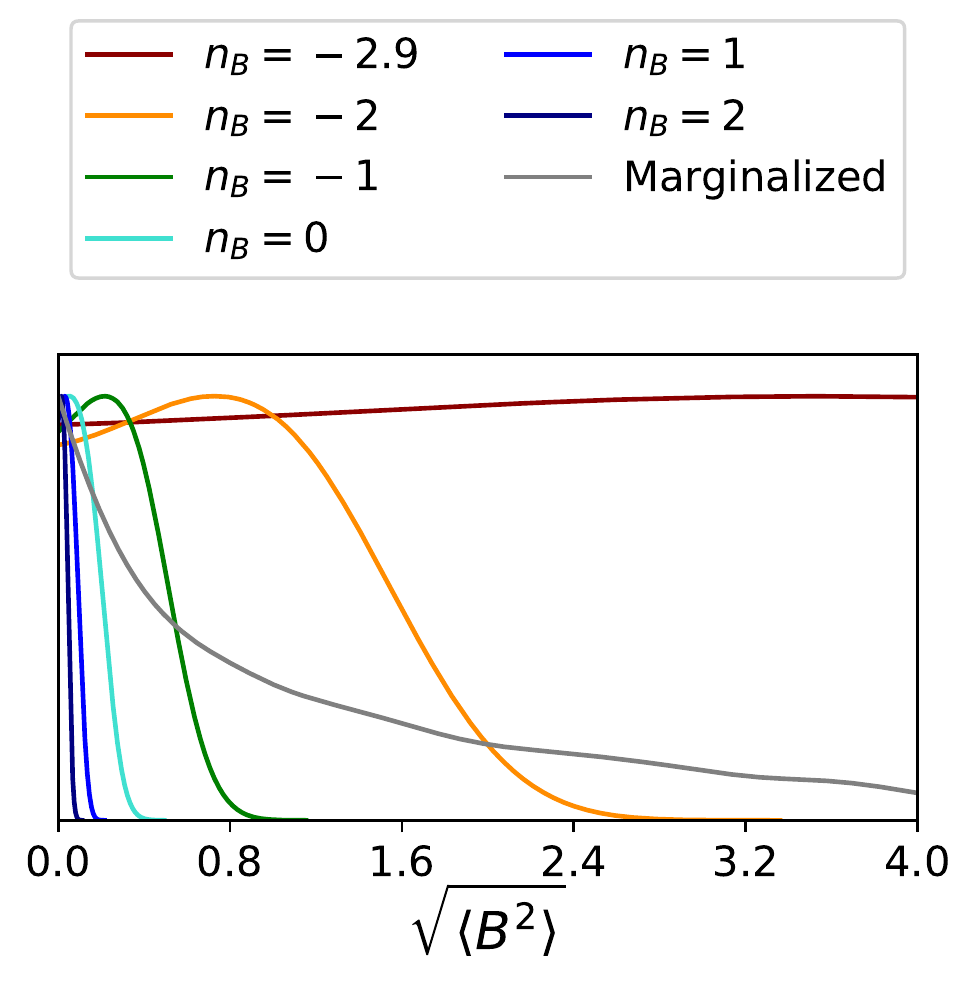}
    \caption{One dimensional posterior distribution for the amplitude of the PMFs for all the spectral indices considered using only the ambipolar effect.}
    \label{fig:1D_AMBI}
\end{figure}
\begin{figure}
    \centering
    \includegraphics[width=0.5\textwidth]{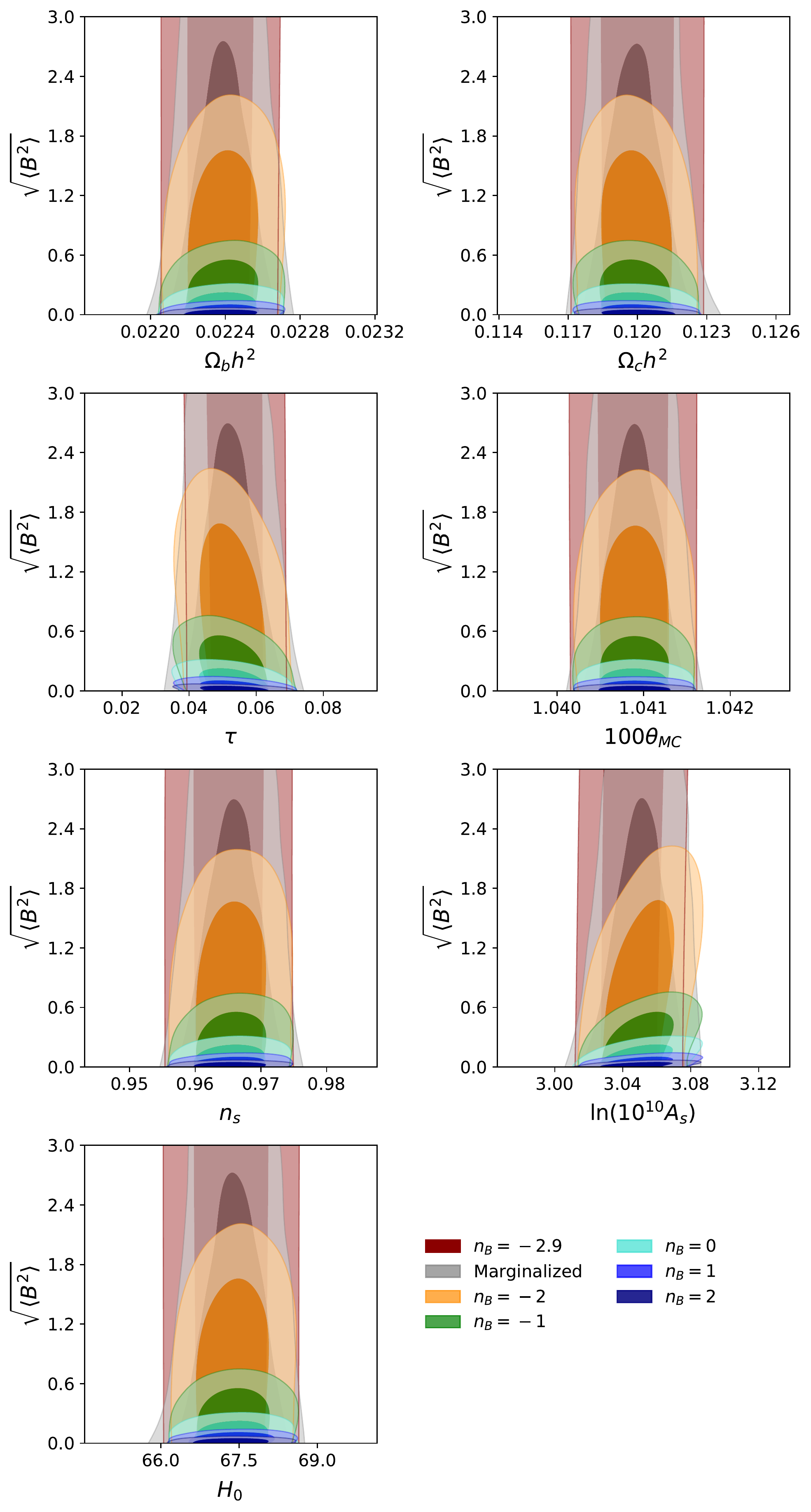}
    \caption{Two dimensional posterior distributions for the standard six cosmological parameters and the amplitude of the PMFs for all the spectral indices considered using only the ambipolar effect.}
    \label{fig:2D_AMBI}
\end{figure}
\begin{table}
\center
\begin{tabular}{|c|c|}
Ambipolar diffusion\\
\hline
$n_{\rm B}$ &$\sqrt{\langle  B^2 \rangle} \, (\mathrm{nG}) $   \\
\hline
2 &  $<0.058$\\
\hline
 1& $<0.12$\\
\hline
0 & $<0.26$\\
\hline
-1 &$<0.62$\\
\hline
-2 & $<1.84$\\
\hline
-2.9 &  $-$($<6.25$ at 68\% C.L.)\\
\hline
[-2.9,2]&  $<3.40$ \\
\hline
\end{tabular}
\caption{\label{tab:1}
 Constraints on the PMF amplitude both for a fixed spectral index and the marginalized case over $n_B$ by using only the ambipolar diffusion effect, constraints are at 95\%C.L.}
\end{table}
\subsection{Ambipolar diffusion}
We start by analysing the constraints on the PMFs amplitude when we consider only the ambipolar diffusion effect.
The posterior distributions for the PMF amplitude are presented in \autoref{fig:1D_AMBI} and in \autoref{tab:1}.
We note the usual trend of the ambipolar diffusion constraints (for both fixed $n_{\mathrm B}$ and the marginalized case) which reflects the dependence of the effect on the angular power spectra with the spectral index. Larger spectral indices are strongly constrained by ambipolar diffusion whereas the negative indices provide looser constraints. The almost scale invariant case is not constrained at 95\,\% C.L. within the large prior of [0,10] nG but provides only a 68\,\% C.L. upper limit.The marginalized constraint is dominated by the lower spectral indices setting at the 3\,nG level.

In \autoref{fig:2D_AMBI} we show the two-dimensional constraints of the amplitude of the fields with the other cosmological parameters from the standard model. Although not strong, we note some degeneracies and in particular with the optical depth and the overall scalar fluctuations amplitude. This is related to the effect on the angular power spectrum as seen in \autoref{fig:AMBI_APS} where on the smaller angular scales we have both in temperature and polarization an increase of the spectrum and on the large scales the impact on the reionization bump in the E-mode polarization which can create confusion with the reionization history.

\begin{figure}
    \centering
    \includegraphics[width=0.5\textwidth]{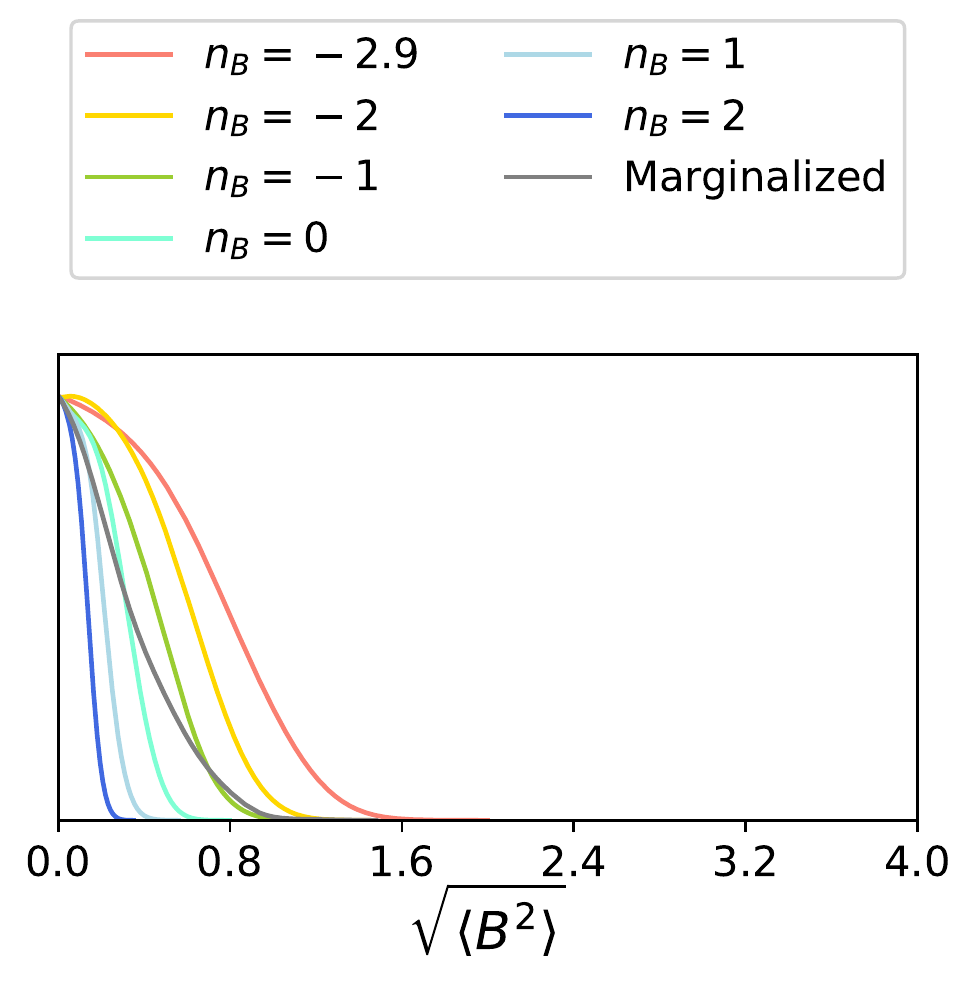}
    \caption{ One dimensional posterior distribution for the amplitude of the PMFs for all the spectral indices considered using only the MHD effect.}
    \label{fig:1D_MHD}
\end{figure}
\begin{table}
\center
\begin{tabular}{|c|c|}
MHD decaying turbulence\\
\hline
$n_{\rm B}$ &$\sqrt{\langle  B^2 \rangle} \, (\mathrm{nG}) $   \\
\hline
2 &  $<0.18$\\
\hline
 1& $<0.27$\\
\hline
0 & $<0.41$\\
\hline
-1 &$<0.63$\\
\hline
-2 & $<0.79$\\
\hline
-2.9 &  $<1.05$\\
\hline
[-2.9,2]&  $<0.68$ \\
\hline
\end{tabular}
\caption{\label{tab:2}
Constraints on the PMF amplitude both for a fixed spectral index and the marginalized case over $n_B$ by using only the MHD decaying turbulence effect, constraints are at 95\%C.L.}
\end{table}
\begin{figure}
    \centering
    \includegraphics[width=0.5\textwidth]{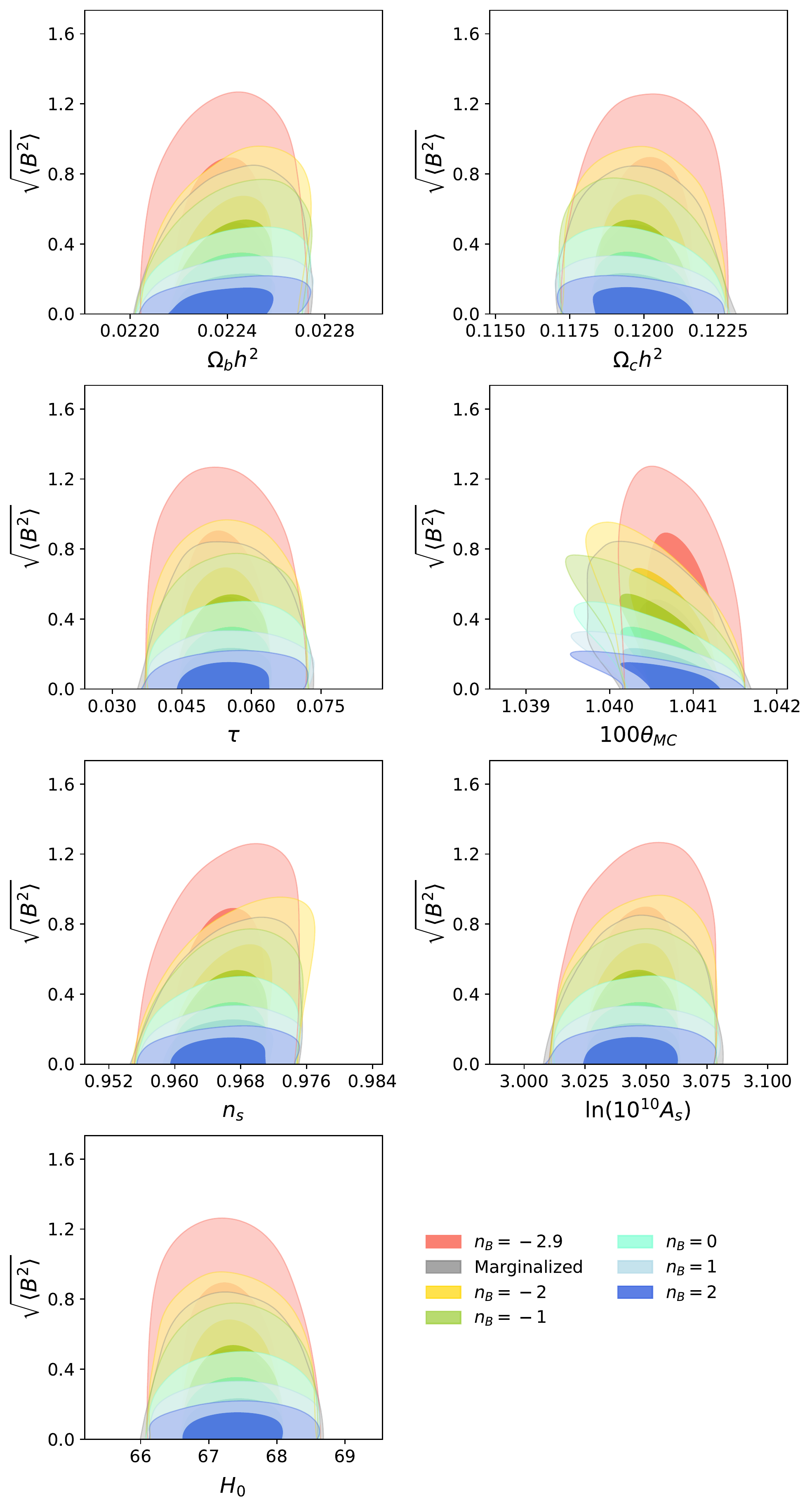}
    \caption{Two dimensional posterior distributions for the standard six cosmological parameters and the amplitude of the PMFs for all the spectral indices considered with only the MHD decaying turbulence effect.}
    \label{fig:MHD_2D}
\end{figure}
\subsection{MHD decaying turbulence}
We now consider only the MHD decaying turbulence effect to constrain the PMFs amplitude. Considering the nature of the effect we expect a more homogeneous constraining power with varying spectral index with respect to the ambipolar diffusion case, with the different indices at similar level of constraints.
In \autoref{fig:1D_MHD} we present the constraints on the PMF amplitude for the different cases considered. The constraints are then presented in \autoref{tab:2}. As expected, the constraints for the various spectral indices and the marginalized case are at the same level just below the nG threshold with only the almost scale invariant case slightly over. The weaker dependence of the constraints on the spectral indices leads to a marginalized constraint perfectly in line with the sub-nG observed in all fixed indices but the almost scale invariant.
In \autoref{fig:MHD_2D} we show the correlation of the PMFs amplitude with the standard parameters of the cosmological model. Also in this case we note some degeneracies but completely different from the ones observed in the ambipolar diffusion case. For the MHD decaying turbulence we have a strong degeneracy of the PMF amplitude with the angular diameter distance to the last scattering, the $\Theta$ parameter, related to the effect of MHD decaying turbulence on the acoustic peaks region of the angular power spectra which affects also the first peak. Another weak degeneracy is found with the scalar spectral index, $n_s$ but in this case the degeneracy looks stronger for intermediate negative indices with the almost scale invariant and the positive indices less affected.

\subsection{Combined effect}
\begin{figure}
    \centering
    \includegraphics[width=0.5\textwidth]{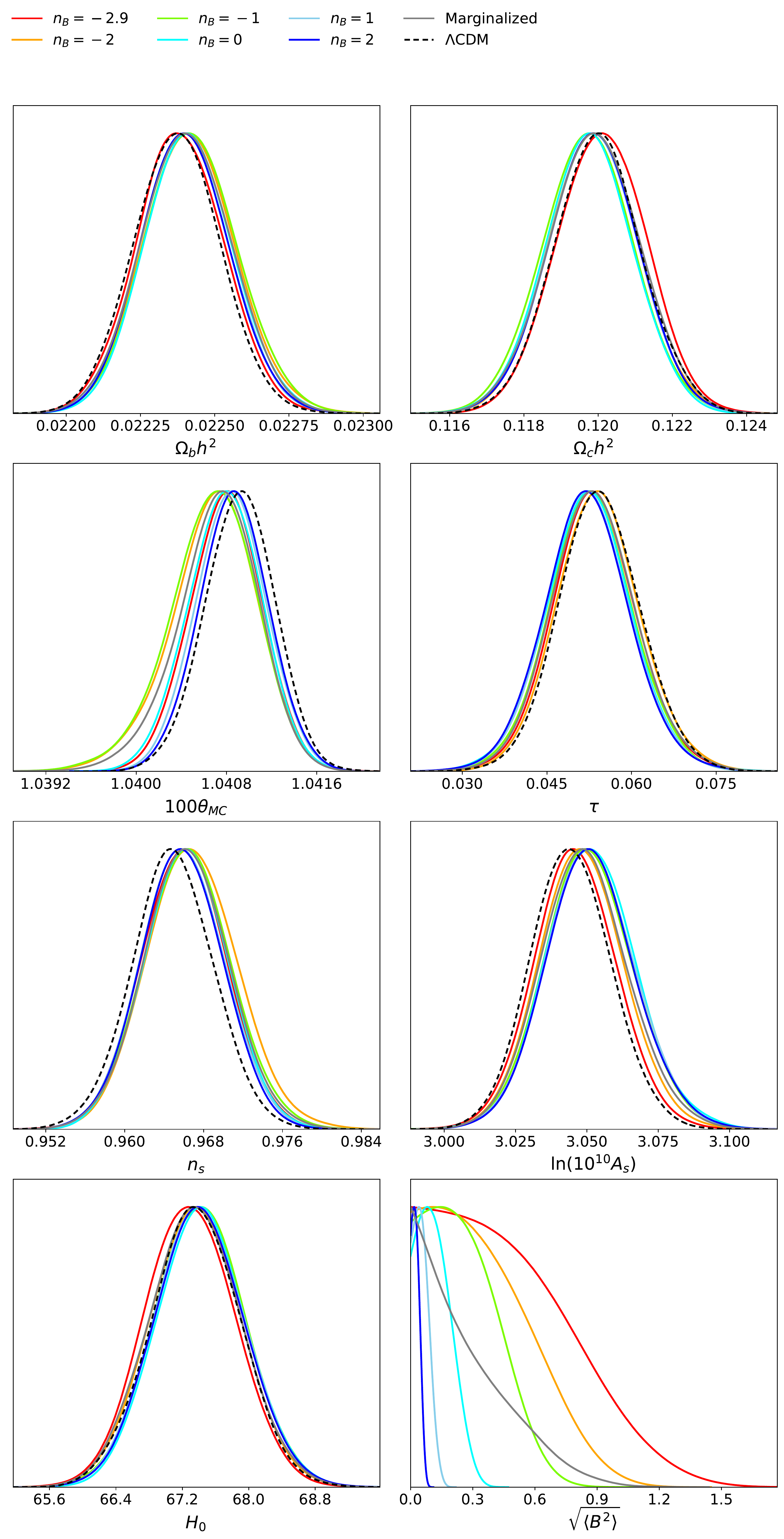}
    \caption{Posterior distributions for the standard six cosmological parameters for all the spectral indices considered compared with the $\Lambda$CDM for the combined case.}
    \label{fig:1D}
\end{figure}

\begin{figure}
    \centering
    \includegraphics[width=0.5\textwidth]{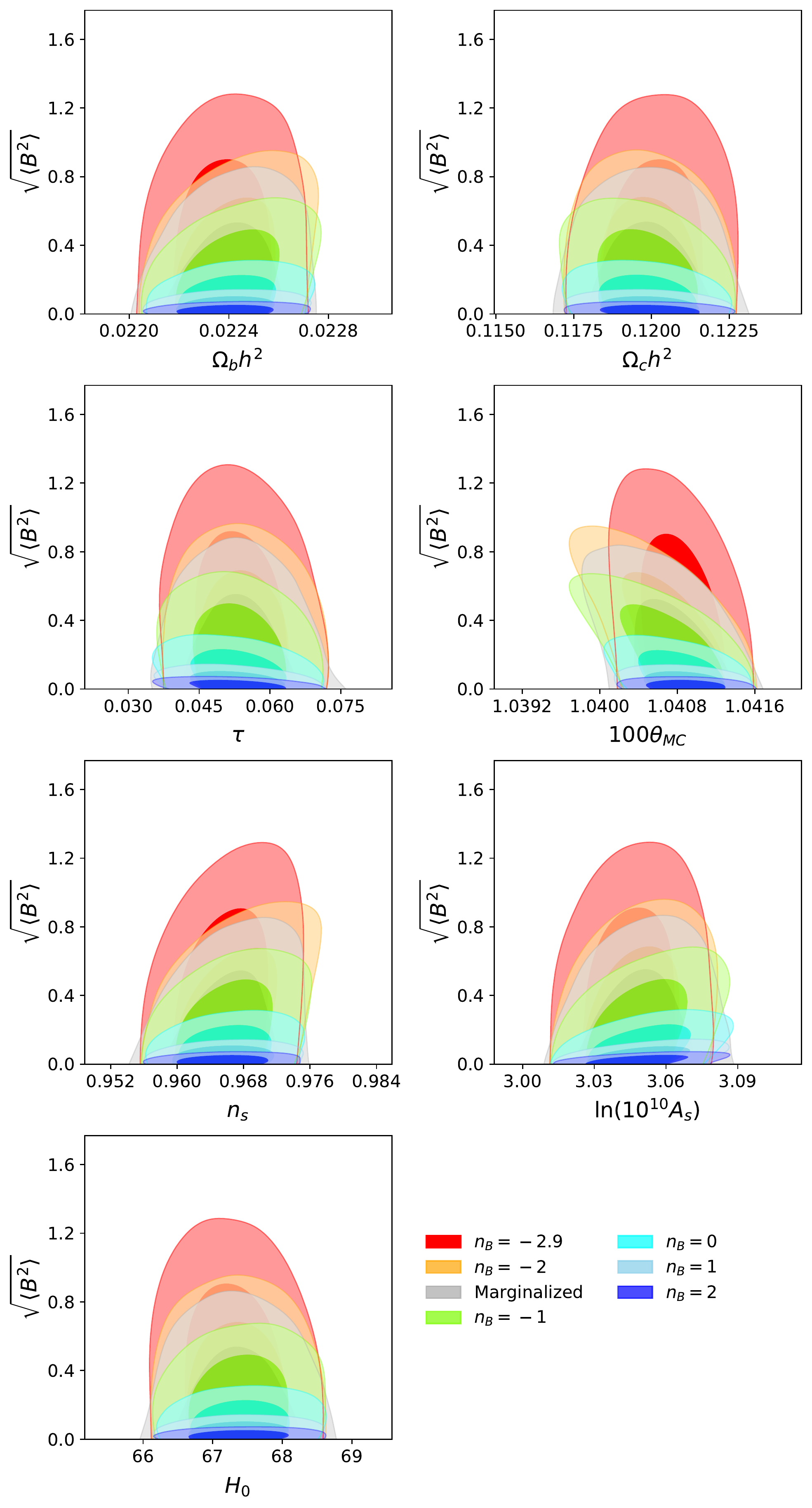}
    \caption{Two dimensional posterior distributions for the standard six cosmological parameters and the amplitude of the PMFs for all the spectral indices considered.}
    \label{fig:2D}
\end{figure}

We now consider the combination of ambipolar diffusion and MHD decaying turbulence and analyse the constraints on PMFs amplitude. The complementarity of the two separate effects with the ambipolar diffusion stronger for positive spectral indices and the MHD decaying turbulence instead homogeneous among different indices allows to put tighter constraints on both ends of the PMFs spectrum. This can be seen from the constraints in \autoref{tab:3} where we we note the influence of both effects with a tight constraints on positive indices and nG level on the almost scale invariant. The marginalized constraint remains sub-nG in the perfect combination of the two effects. One and two dimensional posteriors for the magnetic and cosmological parameters are shown in \autoref{fig:1D} and \autoref{fig:2D}. For this combined case we compare the parameters of the standard model to the ones from the $\Lambda$CDM {\it Planck} 2018 baseline case. In particular, in \autoref{fig:1D} we note how the contribution of PMFs causes a significant shift in at least three of the main cosmological parameters with the mostly affected being the angular diameter distance to last scattering and the scalar fluctuations primordial amplitude. Minorly affected are also the scalar spectral index and the optical depth. For the optical depth and scalar amplitude we note that the main shift is provided by the positive spectral indices because those shift are caused by the ambipolar diffusion contribution whose effect as we shown is stronger for positive indices. For the angular diameter distance and the scalar spectral index instead the main driver is the MHD decaying turbulence and we observe a stronger deviation for the intermediate indices as observed for the MHD case alone.
The shifts in the parameters are reflected by the two dimensional posterior distribution where the degeneracies of the amplitude of PMF with the parameters [$\Theta, n_s, A_s, \tau$] are evident.
\begin{table}
\center
\begin{tabular}{|c|c|}
Combined effect\\
\hline
$n_{\rm B}$ &$\sqrt{\langle  B^2 \rangle} \, (\mathrm{nG}) $   \\
\hline
2 &  $<0.06$\\
\hline
 1& $<0.12$\\
\hline
0 & $<0.26$\\
\hline
-1 &$<0.56$\\
\hline
-2 & $<0.79$\\
\hline
-2.9 &  $<1.06$\\
\hline
[-2.9,2]&  $<0.69$ \\
\hline
\end{tabular}
\caption{\label{tab:3}
Constraints on the PMF amplitude both for a fixed spectral index and the marginalized case over $n_B$ by using the combination of both ambipolar diffusion and MHD decaying turbulence, constraints are at 95\%C.L.}
\end{table}
\subsection{Alternative low-$\ell$ polarization likelihood}

For the combined effect case we also study the alternative low-$\ell$ polarization likelihood \texttt{SROLL2}, based on  contamination levels \citep{Delouis:2019bub}. Within $\Lambda$CDM the use of this likelihood increases the central value of the integrated optical depth and reduces the error bars \citep{Pagano:2019tci}. We therefore test its importance for the PMFs heating model.  

\begin{figure}
    \includegraphics[width=0.25\textwidth]{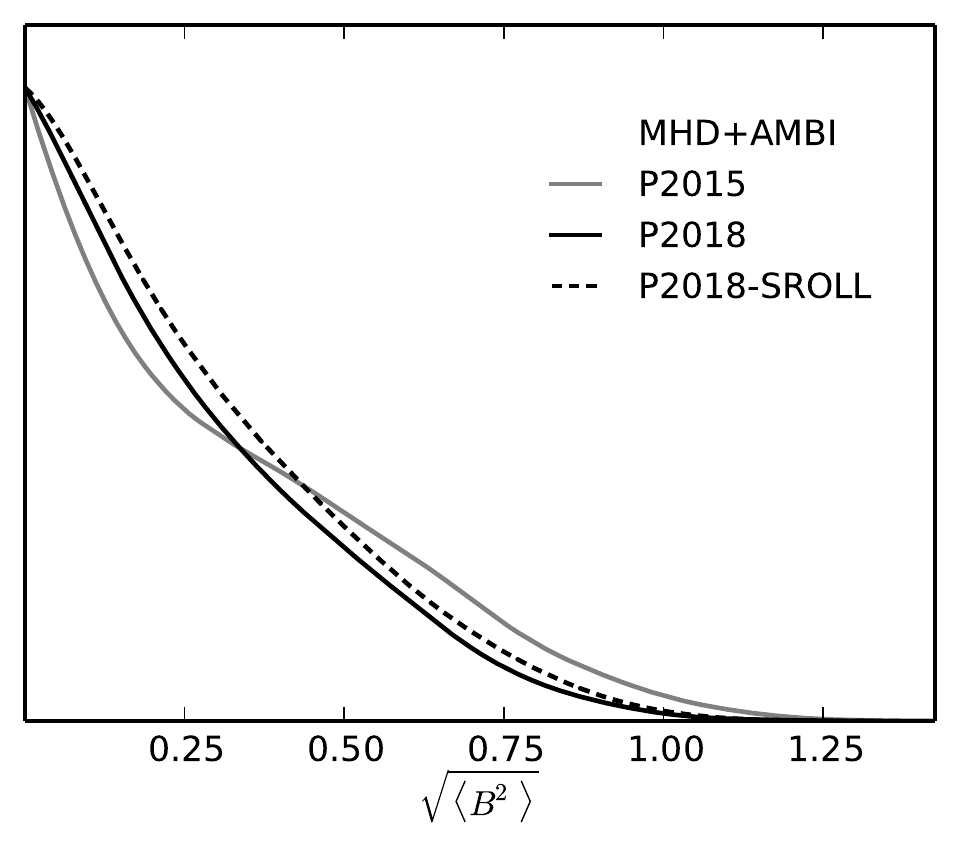}\includegraphics[width=0.25\textwidth]{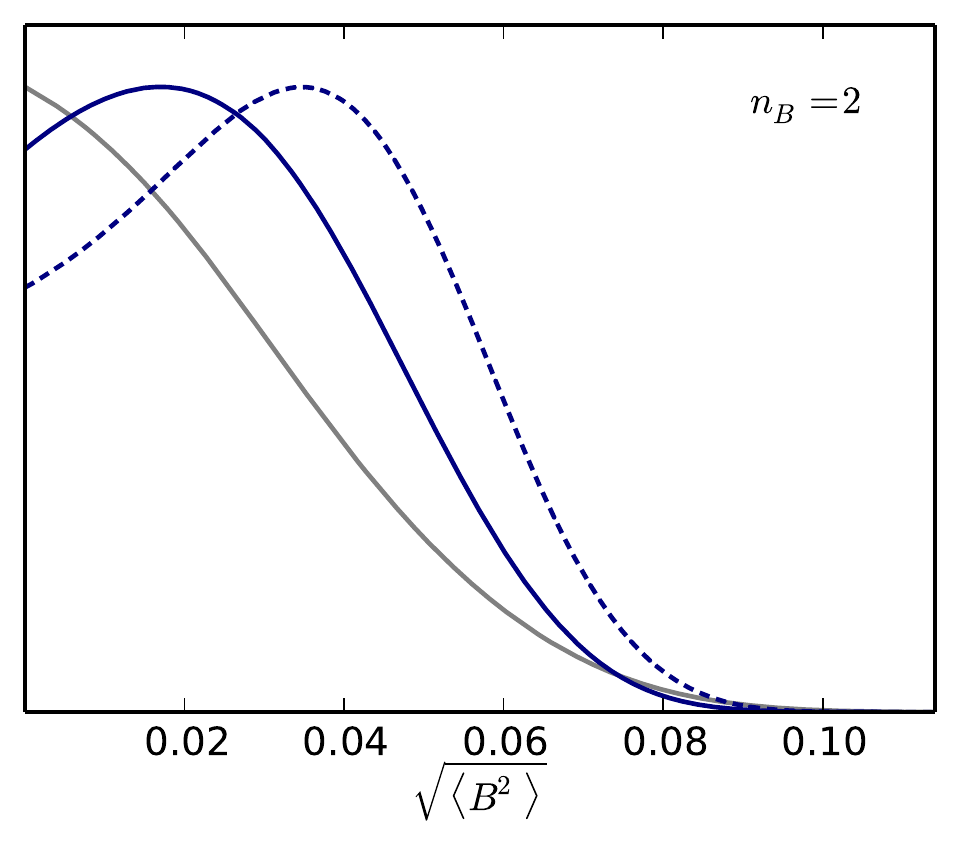}\\\includegraphics[width=0.25\textwidth]{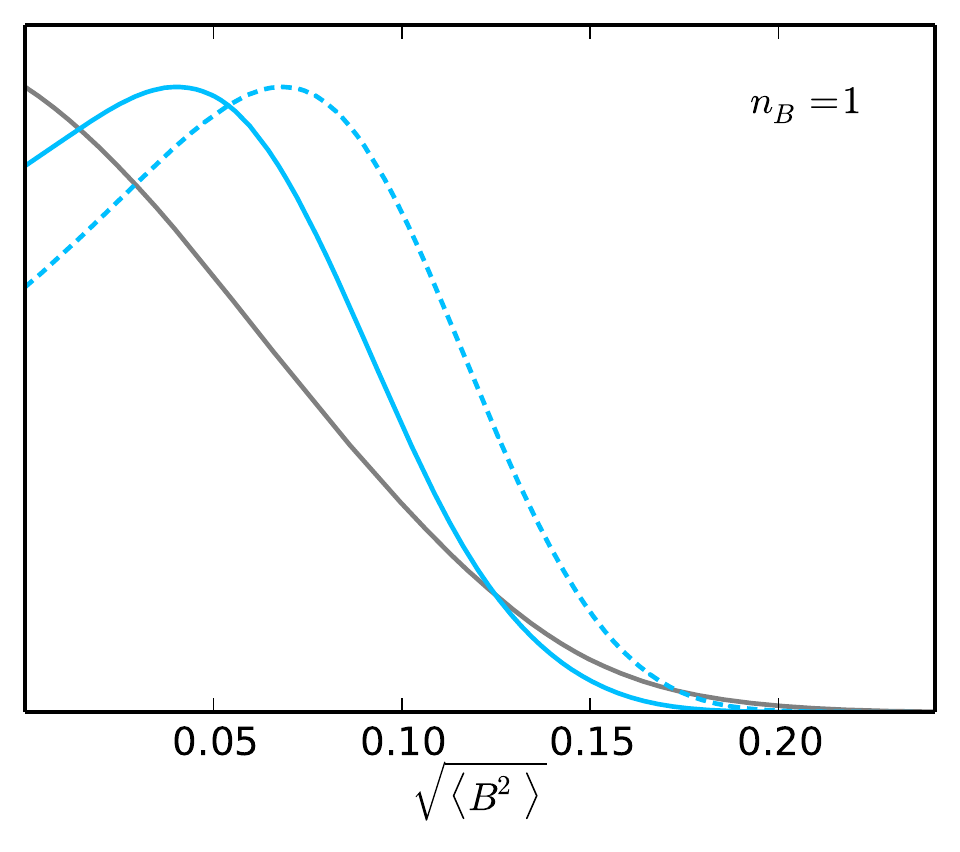}\includegraphics[width=0.25\textwidth]{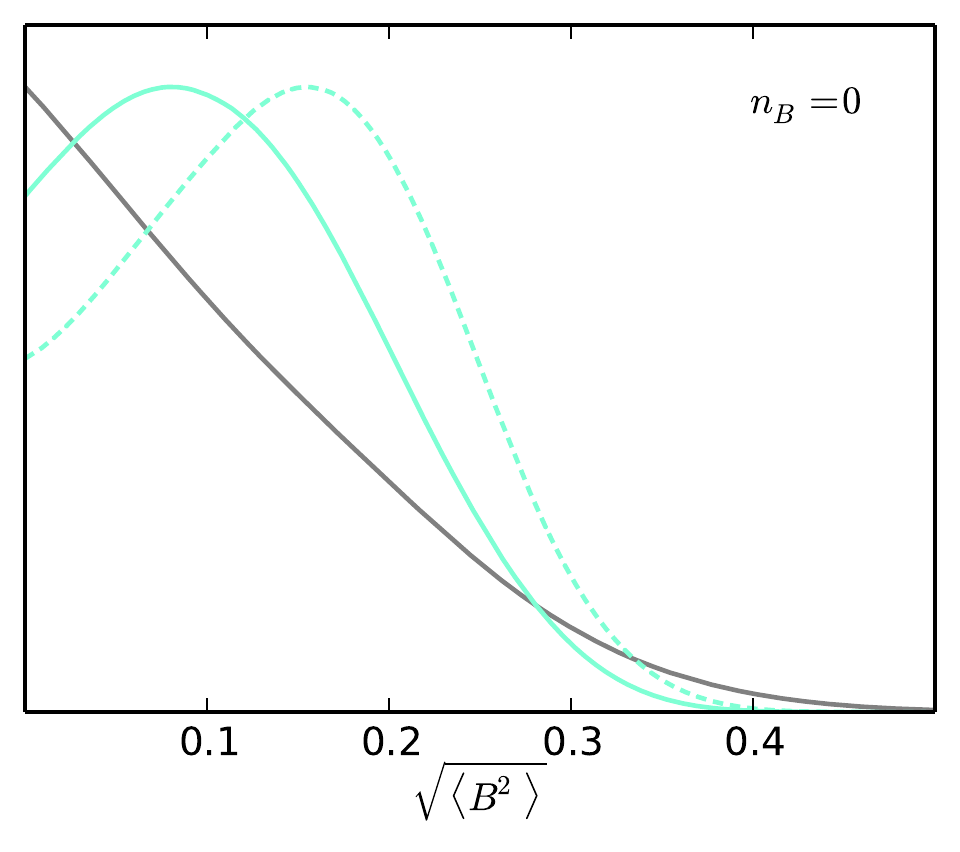}\\\includegraphics[width=0.25\textwidth]{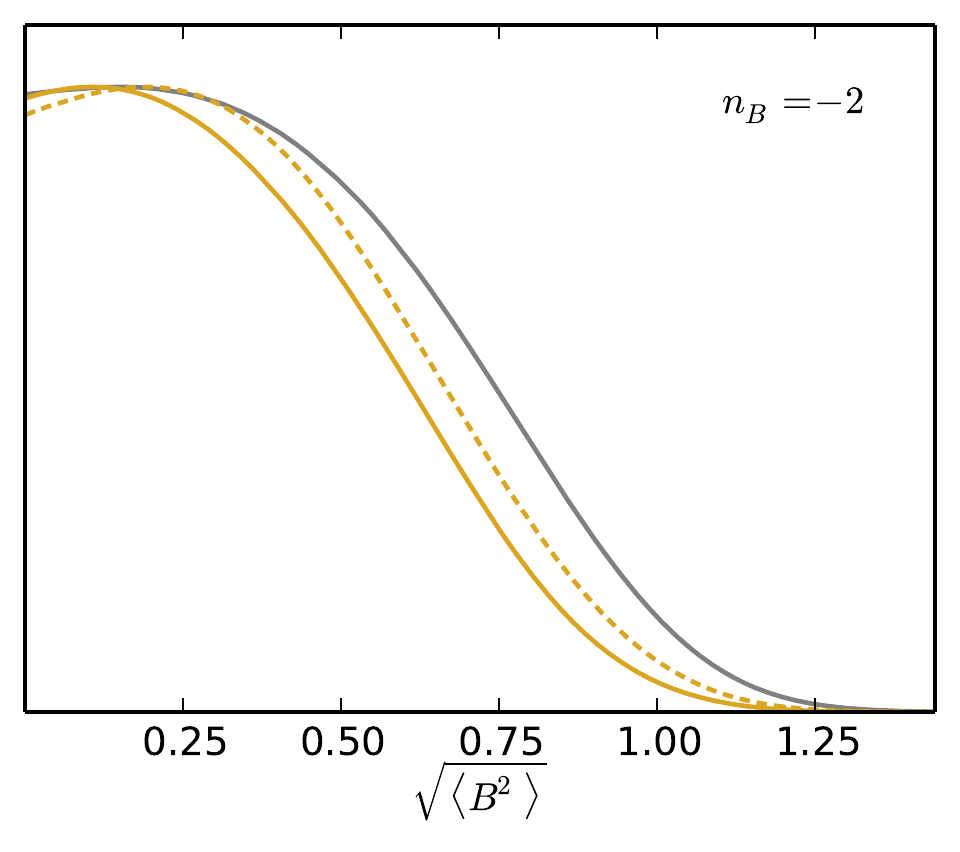}\includegraphics[width=0.25\textwidth]{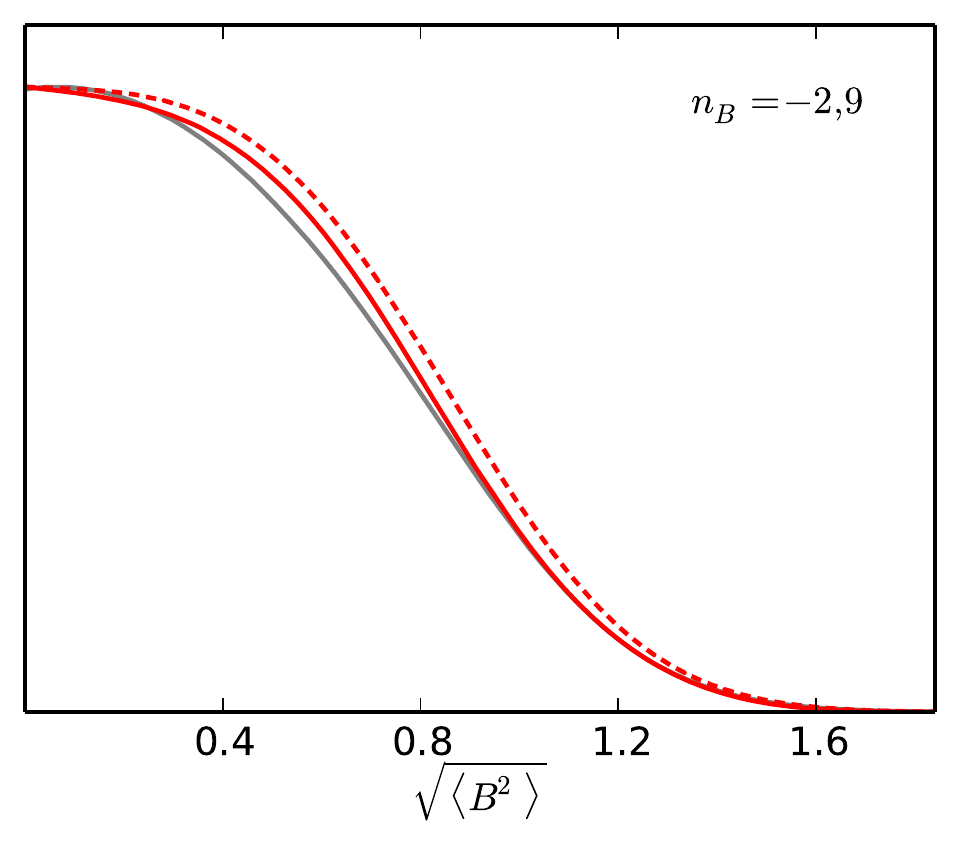}
    \caption{Constraints on the amplitude of PMFs using the combination of ambipolar diffusion and MHD turbulence. The different curves, solid colored, dashed colored and thin black represent {\it Planck} 2018 baseline, {\it Planck} 2018+\texttt{SROLL2} and {\it Planck} 2015}
    \label{fig:COMBO}
\end{figure}

The constraints on PMFs amplitude are show in \autoref{tab:4} where we report both the 95\% C.L. upper bound but also the 68\% C.L. in parentheses. We report also the 1$\sigma$ results because the change from the {\it Planck} baseline to the \texttt{SROLL2} likelihood causes a 68\% C.L. detection of the PMF amplitude for the blue spectral indices (which was marginally present only for the $n_{\mathrm B}=0$ case for {\it Planck} baseline). The change in the posterior is clearly visible in \autoref{fig:COMBO} where we compare the one dimensional posterior distributions from the {\it Planck} 2018 baseline with {\it Planck} 2015 and {\it Planck} 2018-\texttt{SROLL2} for the different spectral indices.
\begin{table}
\center
\begin{tabular}{|c|c|}
Combined effect-\texttt{SROLL2}\\
\hline
$n_{\rm B}$ &$\sqrt{\langle  B^2 \rangle} \, (\mathrm{nG}) 95\% C.L. (68\% C.L.)$   \\
\hline
2 &  $<0.07\,(0.03_{-0.02}^{+0.02})$\\
\hline
 1& $<0.14\,(0.07_{-0.05}^{+0.03})$\\
\hline
0 & $<0.28\,(0.15_{-0.09}^{+0.08})$\\
\hline
-2 & $<0.84\,(<0.49)$\\
\hline
-2.9 &  $<1.08\,(<0.63)$\\
\hline
[-2.9,2]&  $<0.72\,(<0.36)$ \\
\hline
\end{tabular}
\caption{\label{tab:4}
Constraints on the PMF amplitude both for a fixed spectral index and the marginalized case over $n_B$ with {\it Planck} 2018+\texttt{SROLL2} by using the combination of both ambipolar diffusion and MHD decaying turbulence, constraints are at 95\%C.L. and in parentheses we report the 68\% C.L..}
\end{table}

The first aspect we note is the dependence of the comparison with the spectral index. Negative indices tend to provide only little evolution of the constraints among the three different datasets from {\it Planck} 2015 to {\it Planck} 2018+\texttt{SROLL2} with the marginalized case almost unmodified. The positive indices and in particular the cases around $n_{\mathrm B}=0$ provide instead a strong evolution,  with the general trend of {\it Planck} 2018 providing larger constraints with respect to the 2015 data strengthen by the \texttt{SROLL2} likelihood. This most recent likelihood for positive indices is still compatible with an upper limit but it shows pronounced peaks in the posteriors corresponding to 68\% C.L. detections.
The reason of this evolution with the data of the constraints is to be found in the changes to the E-mode polarization between the different releases. In addition, being the positive indices most affected we must search in the ambipolar diffusion effect on large angular scales EE. On these scales the {\it Planck} power spectra has significantly changed among the releases with the 2015 data more compatible with high redshift reionization tails (see for example \cite{Hazra:2017gtx})  and higher optical depth, whereas {\it Planck} 2018 have reduced the duration of reionization and the optical depth making the power spectra more compatible with the change in shape of the reionization bump provided by the ambipolar diffusion. For this reason we observe a large allowed amplitude for the fields, the reionization bump shape produced by the ambipolar diffusion is in a better agreement with {\it Planck} 2018 data and even more with the reduced error bars of \texttt{SROLL2} allowing for a larger PMFs amplitude and even a tentative 1-$\sigma$ detection.  In \autoref{tab:5} we show the comparison of the different constraints for the combined case.
\begin{table}
\center
\begin{tabular}{|c|c|c|c|}
\hline
$n_{\rm B}$ & &$\sqrt{\langle  B^2 \rangle} \, (\mathrm{nG}) $ \\
\hline
& {\it Planck} 2015 & {\it Planck} 2018 & {\it Planck} 2018-\texttt{SROLL2} \\
\hline
2 & $<0.06$ &  $<0.06$ & $<0.07$\\
\hline
 1 &$<0.13$ & $<0.12$ & $<0.14$\\
\hline
0 & $<0.30$ & $<0.26$ & $<0.28$ \\
\hline
-2 & $<0.90$ & $<0.79$ & $<0.84$\\
\hline
-2.9 & $<1.06$ & $<1.06$ & $<1.08$\\
\hline
[-2.9,2] & $<0.83$ & $<0.69$ & $<0.72$\\
\hline
\end{tabular}
\caption{\label{tab:5}
Comparison of the constraints from the combined effects for different datasets, from {\it Planck} 2015 in the first column to {\it Planck} 2018 baseline in the second column and {\it Planck} 2018-\texttt{SROLL2} in the third column.}
\end{table}

\subsection{Damping scale}

As already stated, the PMF post-recombination heating by ambipolar diffusion depends critically on the physics at 
the damping scale \cite{Paoletti:2018uic}. In the following we further explore this dependence by allowing $k_D$ to vary instead of keeping it fixed \citep{Subramanian1998}.  

The results for the PMF amplitude and the damping scale are presented in \autoref{tab:6}. 
We note that by allowing the damping scale to vary, the upper bounds on the PMF slightly decrease but are well degenerate with the damping scale. The additional degree of freedom almost erases the dependence of the constraints on the spectral index levelling all the constraints at the level of $0.3$--$0.5$\,nG. This is related to the decreased dependence on the spectral index of the ambipolar diffusion, opening the damping scale removes part of this dependence leaving the constraints more or less at the same level. The damping scale constraints vary with a slightly larger damping wavenumber for the negative spectral indices which have also slightly larger upper bounds. This results indicate that in the future a detailed study of the importance of the physics of the damping will be needed for effects as the ambipolar diffusion.
\begin{table}
\center
\begin{tabular}{|c|c|c}
\hline
$n_{\rm B}$  &$\sqrt{\langle  B^2 \rangle} \, (\mathrm{nG}) $  & $10^5 k_D\sqrt{\langle  B^2 \rangle} \, (\mathrm{Mpc}^{-1} \mathrm{nG})$\\
\hline
2 & $<0.38$ &  $<0.56$ \\
\hline
1 &$<0.33$ & $<0.61$ \\
\hline
0 & $<0.35$ & $<0.59$ \\
\hline
-2 & $<0.39$ & $<0.86$\\
\hline
-2.9 & $<0.47$ & $<2.0$\\
\hline
[-2.9,2] & $<0.42$ & $<0.53$\\
\hline
\end{tabular}
\caption{\label{tab:6}
95 \% CL constraints for the case where we leave the damping scale as a parameter free to vary.}
\end{table}

\subsection{Possible confusion with the lensing amplitude}
\begin{figure*}
    \centering
    \includegraphics[width=\textwidth]{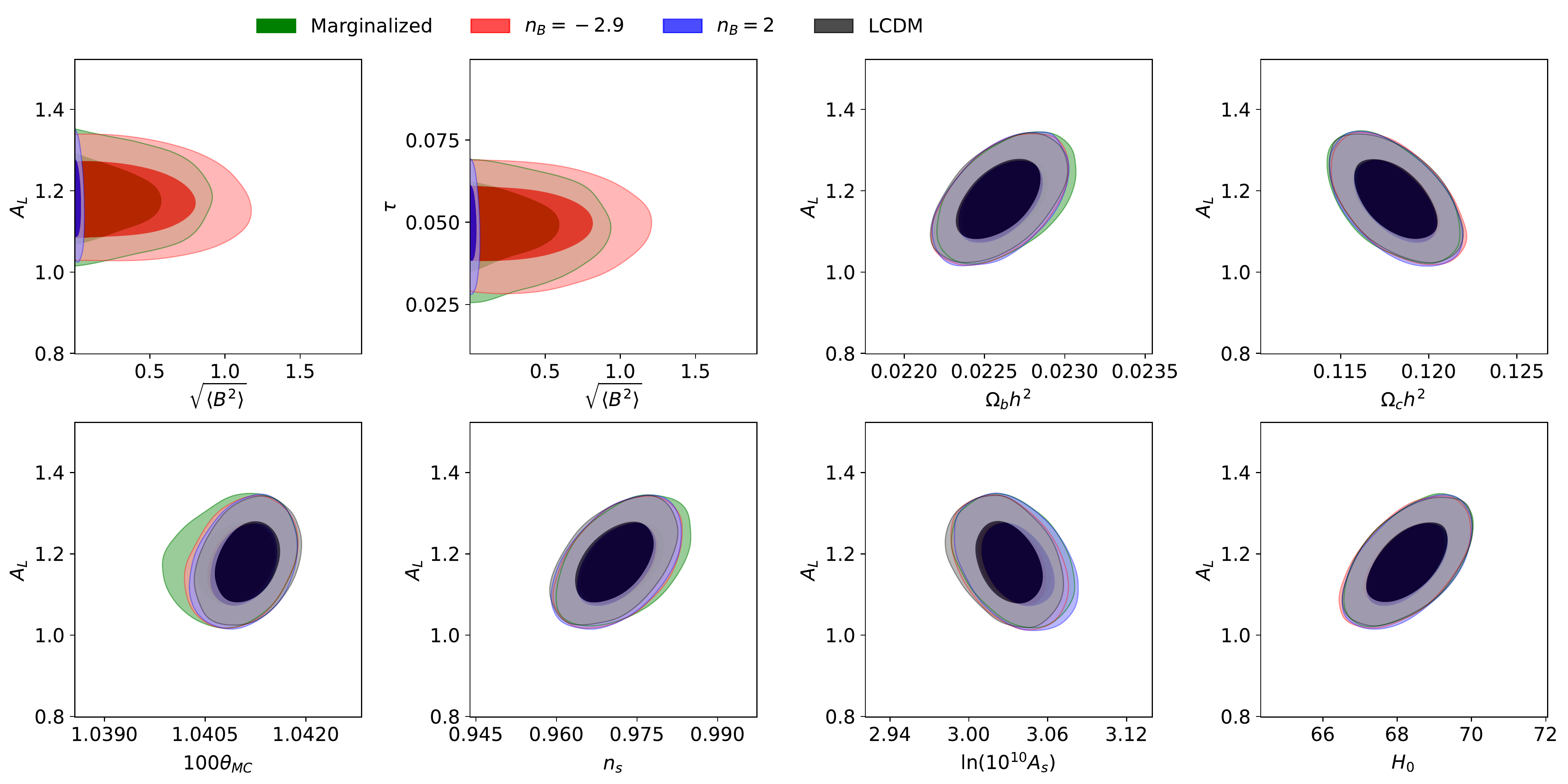}
    \caption{Constraints on PMF and lensing amplitudes for the combination of ambipolar diffusion and MHD turbulence. We do not consider the lensing likelihood for this analysis}
    \label{fig:alens}
\end{figure*}

The peculiar nature of the heating effects causes a change in the acoustic peaks region of the power spectrum and in particular we wanted to investigate if this effect can be somehow correlated with the effect of the gravitational lensing in the angular power spectrum. One of the well known curiosities of the {\it Planck} 2018 results is the slightly larger amplitude of the lensing effect on the angular power spectrum with respect to its value estimated from the non-Gaussianity induced \citep{Planck:2018vyg}. We performed three analyses, considering only the marginalized case and the two extrema of the spectral index range, with the amplitude of the lensing as an additional free parameter. For this case we do not add the lensing likelihood to the {\it Planck} 2018 data since it is based on the non-Gaussianity induced by the lensing and would bias the results towards the $A_L=1$.
The results are shown in \autoref{fig:alens} compared with the standard $\Lambda CDM$ case. We note that there are no significant correlations between the PMF and lensing amplitude with the PMFs providing $A_L=1.175\pm 0.067$, $A_L=1.177\pm 0.066$ and $A_L=1.180_{-0.066}^{+0.065}$ respectively for the $n_{\mathrm B}=2$, $n_{\mathrm B}=-2.9$ and marginalized cases to be compared with the {\it Planck} result  $A_{L}=1.180\pm 0.065$~\citep{Planck:2018vyg}. The contribution of the PMFs only brings a minor reduction of the central value of a fraction of sigma. 

\section{Conclusions}

Primordial magnetic fields can be the progenitors of the cosmic magnetism or at least may have played a significant role in its generation.  
If they were to be present then they would have had an unavoidable impact on the whole Universe history and should have left their footprints on CMB anisotropies. We have reviewed how their dissipation after recombination strongly impacts E-mode polarization and represents one of the best avenues to investigate and constrain PMF characteristics. 

The {\it Planck} 2018 release and subsequent updates have improved the E-mode polarization calling for an update of our previous results \citep{2015MNRAS.451.2244C,Paoletti:2018uic}.
Our results show some improvements especially for some spectral indices, but not for all configurations of the PMF with in particular the almost scale invariant case slightly degraded. This is mainly traceable to two factors. The first is the change of large scale polarization likelihood approach between 2015 and 2018 {\it Planck} releases that in 2018 adopted a simulation based likelihood without TE cross-correlation which can affect the constraints of models as the heating by PMF. The second is the change occurred in the data. The new E-mode polarization has a spectral shape which is better fitted by the contribution on large angular scales of PMFs allowing for larger PMF amplitudes with respect to {\it Planck} 2015 and some marginal 68\,\% C.L. detection especially when SROLL 2 reanalysis of the data is used. The difference in the spectral indices dependence of the two effects is again reflected by the constraints with a flat behaviour of the MHD decaying turbulence and instead a stronger constraints for blue indices provided by the ambipolar diffusion. The overall constraining power stands at the nG level with considerably sub-nG levels only for positive spectral indices.\\

We have then investigated the importance of keeping fixed the damping scale in our analysis \citep{Subramanian1998}.
The major results of co-sampling the damping scale value is the loss of dependence on the spectral index and an increased difficulty in the convergence of the MCMC. The resulting constraints on the amplitude of the fields become flat at the level of half nG and the damping scale is only poorly constrained. These results show at the same time the robustness of our constraints on the amplitude of PMFs and the necessity for further improvements in the understanding of the physics of the damping scale.

The impact of PMF dissipation on the angular power spectra of CMB anisotropies can create confusion with the effect of changing the standard cosmological parameter. For the first time we have investigated the degeneracies of the PMF parameters with the parameters of the cosmological model which are co-varied in our approach. The double leverage of the effects at both large and small angular scales makes ambipolar diffusion little degenerate with the undelying cosmological model with the only degeneracy being with the optical depth and the scalar amplitude, as expected due to the impact on the reionization bump in the E-mode polarization.
On the other hand, due to its nature, the MHD decaying turbulence effect impacts mainly the acoustic region peaks in both shifts and amplitudes of the oscillations, causing degeneracies with parameters as the angular diameter distance, which is fixed by the first peak position, and the scalar spectral index. The presence of these degeneracies implies having both large and small angular scale leverages in the data and therefore outlines these models as one of the best target for future data. Future combined data from {\it LiteBIRD} on large angular scales \citep{Sugai:2020pjw,LiteBIRD:2020khw} and ground based experiments on the small ones \citep{SimonsObservatory:2018koc,Abazajian:2019eic} promise to provide a cosmic variance limited E-mode polarization measurement on all the scales relevant for the CMB, representing a really favourable scenario for the PMF effect on the ionization history.

\small
\section*{Acknowledgments}
DP and FF acknowledge financial support by the agreement n. 2020-9-HH.0 ASI-UniRM2 ``Partecipazione italiana alla fase A della missione LiteBIRD".
DP acknowledges the computing centre  of Cineca and INAF, under the coordination of the ``Accordo Quadro MoU per lo svolgimento di attività congiunta di ricerca Nuove frontiere in Astrofisica: HPC e Data Exploration di nuova generazione", for the availability of computing resources and support with the project INA17-C5A42. DP also acknowledges the use of Cineca under the INFN agreement for InDark.
JC was supported by the Royal Society as a Royal Society University Research Fellow at the University of Manchester, UK (No.~URF/R/191023)and the ERC Consolidator Grant {\it CMBSPEC} (No.~725456).
JARM acknowledges financial support from the Spanish Ministry of Science and Innovation (MICINN) under the projects AYA2017-84185-P and PID2020-120514GB-I00.

{
\vspace{-0mm}
\small
\bibliographystyle{mn2e}
\bibliography{Lit}
}

\end{document}

%% file: Befehle.tex
\newcommand{\Mpc}{{\rm Mpc}}

\newcommand{\id}{{\,\rm d}}

\newcommand{\beq}{\begin{equation}}   %

\newcommand{\eeq}{\end{equation}}   %

\newcommand{\beqa}{\begin{eqnarray}}   %

\newcommand{\eeqa}{\end{eqnarray}}   %

\newcommand{\beal}{\begin{align}}

\newcommand{\enal}{\end{align}}

\newcommand{\bspl}{\begin{split}}

\newcommand{\espl}{\end{split}}

\newcommand{\bsub}{\begin{subequations}}

\newcommand{\esub}{\end{subequations}}

\newcommand{\bmulti}{\begin{multline}}   %

\newcommand{\beqm}{\begin{mathletters}}   %

\newcommand{\eeqm}{\end{mathletters}}   %

\newcommand{\me}{m_{\rm e}}

\newcommand{\Ne}{N_{\rm e}}

\newcommand{\Te}{T_{\rm e}}

\newcommand{\Tg}{T_{\gamma}}

\newcommand{\sigT}{\sigma_{\rm T}}

\newcommand{\pot}[2]{#1 \times 10^{#2}}

\newcommand{\Yp}{Y_{\rm p}}
